\documentclass[prd,showpacs, nofootinbib, floatfix,preprintnumbers]{revtex4}%

\usepackage{amssymb,hyperref, amsmath}
\usepackage[dvips]{color}
\usepackage[dvips]{graphicx}
\usepackage{epsfig}
\preprint{JLAB-THY-08-806}

\begin{document}
\newcommand{\beq}{\begin{equation}}
\newcommand{\eeq}{\end{equation}}
\newcommand{\tr}{\mbox{tr}\,}

\bibliographystyle{apsrev}
\title{Tuning for Three-flavors of Anisotropic Clover Fermions with Stout-link Smearing}

\author{Robert G. Edwards}
\email{edwards@jlab.org} \affiliation{Thomas Jefferson National
Accelerator Facility, Newport News, VA 23606}

\author{B\'alint Jo\'o}
\email{bjoo@jlab.org}
\affiliation{Thomas Jefferson National Accelerator Facility,
Newport News, VA 23606}

\author{Huey-Wen Lin}
\email{hwlin@jlab.org} \affiliation{Thomas Jefferson National
Accelerator Facility, Newport News, VA 23606}

\date{Mar. 26, 2008}

\begin{abstract}
In this work we perform the parameter tuning of three flavors of
dynamical clover quarks on anisotropic lattices. The fermion action
uses three-dimensional spatial stout-link smearing.  The gauge
anisotropy is determined in a small box with Schr\"odinger background
using Wilson-loop ratios. The fermion anisotropy is obtained from
studying the meson dispersion relation with antiperiodic
boundary conditions in the time direction. The spatial and temporal clover coefficients are fixed to the tree-level tadpole-improved values, and we demonstrate that they satisfy the nonperturbative conditions as determined by the Schr\"odinger-functional method. For the desired lattice spacing
$a_s\approx 0.12$~fm and renormalized anisotropy $\xi=3.5$, we find
the gauge and fermionic anisotropies can be fixed to quark mass
independent values up through the strange quark mass. This work lays
the foundation needed for further studies of the excited-state hadron
spectrum.
\end{abstract}

\maketitle
\section{Introduction}

Lattice quantum chromodynamics (QCD) has successfully calculated many
properties of the hadronic spectrum. However, there remain many
challenges for the lattice community to resolve in determining the
myriad states present in QCD. The case of the nucleon spectrum is one
such example. Consider the lowest three states in the $N$ spectrum
($N$, $N^\prime$ ($P_{11}$) and $N^*$ ($S_{11})$), for example. Many
earlier quenched lattice QCD
calculations~\cite{Sasaki:2001nf,Guadagnoli:2004wm,Leinweber:2004it,Sasaki:2005ap,Sasaki:2005ug,Burch:2006cc}
find a spectrum inverted with respect to experiment, with the $N^\prime$
heavier than the opposite-parity state $N^*$.  Although the Kentucky
group~\cite{Mathur:2003zf} managed to find the correct mass ordering
around a pion mass of 300--400~MeV (after taking care of the effects
of the quenched ``ghosts''), no other lattice group has been able to
reproduce the experimental ordering using different
approaches. Furthermore, these are just the lowest few states in the
$N$ spectrum. There are many more states seen in experiment for which
lattice calculations could help in the identification of particle properties.

This situation suggests an urgent need for full-QCD simulations that
can resolve some of these issues. In order to improve the signal of
the excited states (especially for the higher-excited nucleon
spectrum), one needs a lattice with a fine temporal inverse lattice
spacing on the order of $6$~GeV. At the same time, we also want to
avoid finite-volume effects. Current dynamical lattice gauge ensembles manage to have a reasonable lattice volume with spatial dimensions of about $3$~fm; however, typically the inverse lattice spacing is about $2$~GeV, which is not fine enough to allow an accurate determination of more than one excited state. One solution to this problem is to generate anisotropic dynamical lattices.

Anisotropic techniques have been widely adopted in lattice
calculations. Anisotropic relativistic heavy quark actions have been used
for charmonium studies~\cite{Chen:2000ej,Okamoto:2001jb}.
Another main application is for calculations, such as
glueballs~\cite{Morningstar:1999rf} and multiple excited-state
extraction~\cite{Lichtl:2006dt,Basak:2007kj,Dudek:2007wv},
where the anisotropic lattice technique has advantages over isotropic lattices due to the finer temporal lattice spacing.  However, there is a worry that uncontrollable quark mass effects, $m a_s$,
might enter into systematic
errors~\cite{Harada:2001ei,Aoki:2001ra,Hashimoto:2003fs} when $m$ (the quark mass) is large and the spatial lattice spacing $a_s$ is about $0.1$~fm. Since we are working in the light-quark limit, this is not a
major concern here. In order to remove possible ${\cal O}(a)$
systematic errors from the action, we tune our anisotropic action to
be as close to the (on-shell) nonperturbatively correct action as
possible.

Previous results on anisotropic lattices include two-flavor
anisotropic dynamical simulations done by CP-PACS~\cite{Umeda:2003pj}
and TrinLat collaboration~\cite{Morrin:2006tf}.  CP-PACS performed the
first dynamical $N_f=2$ anisotropic tuning of dynamical clover
fermions~\cite{Umeda:2003pj} (without gauge-link smearing), using
the Iwasaki gauge and Sheikholeslami-Wohlert (also called
clover~\cite{Sheikholeslami:1985ij}) fermion actions.
In that study, they set the coefficient of the clover term within the
clover action to tadpole-improved tree-level
values. The TrinLat collaboration~\cite{Morrin:2006tf}
used a two-plaquette Symanzik-improved gauge action with tree-level
tadpole improvement and a Wilson fermion action with a Hamber-Wu
term. One should also note that they adopted stout-link
smearing~\cite{Morningstar:2003gk} of the spatial gauge fields in the
clover action. That is, the gauge fields entering the fermion action
were not smeared in the time direction, preserving the positivity of
the fermion transfer matrix. Only two iterations of stout smearing
were used with a staple weight $\rho=0.22$.

In this work, we will use a three-flavor clover action with
stout-link smearing (in the spatial directions only), and an
$O(a^2)$-improved Symanzik gauge action.  Working in the Schr\"odinger-functional
scheme~\cite{Luscher:1992an,Luscher:1996sc,Luscher:1996ug,Klassen:1997jf},
we determine the gauge anisotropy by computing Wilson loop ratios with
the background field applied in the $z$ direction.  The fermion
anisotropy is determined from the conventional meson dispersion
relation with periodic boundaries in the spatial directions and antiperiodic boundaries in the time direction. The coefficients of the
gauge action are set to the tree-level tadpole-improved values, and the
clover coefficients are fixed at the tree-level stout-link smeared
tadpole-improved values, with the tadpole factors set from numerical
simulation. We demonstrate that the clover coefficients are consistent
with nonperturbative values determined in the Schr\"odinger-functional
scheme. Our configurations have been generated using the
Chroma~\cite{Edwards:2004sx} HMC code with RHMC for all three flavors
and multi-timescale integration. A preliminary study can be found in
Ref.~\cite{Lin:2007yf}.

The structure of this paper is as follows: In Sec.~\ref{sec:Setup}, we
will discuss the details of the actions used in this work, the
stout-link smearing and Schr\"odinger-functional scheme calculations,
and how we determine the coefficients. Then we will cover the
Rational Hybrid Monte Carlo (RHMC) used in this work, how we apply it
on anisotropic lattices with even-odd preconditioning, and how to use
these techniques with stout-link smearing in Sec.~\ref{sec:Algorithm}. We present numerical results in Sec.~\ref{sec:Num}, where the gauge and fermion anisotropy, and PCAC mass are measured, and their
corresponding (tuned) bare values are determined. Some conclusions and
future outlook are presented in Sec.~\ref{Sec:Conclusion}.

\section{Methodology and Setup}\label{sec:Setup}

\subsection{Action}

In this section, we describe the gauge and fermion actions used in
this calculation. For the gauge sector, we use a Symanzik-improved
action which was used in the glueball study of Ref.~\cite{Morningstar:1999rf}. With tree-level tadpole-improved
coefficients, the action is
\begin{eqnarray}\label{eq:aniso_syzG}
S_G^{\xi}[U] &=& \frac{\beta}{N_c\xi_0} \left\{
\sum_{x, s>s^\prime} \left[  \frac{5}{3 u_s^4}{\cal P}_{ss^\prime}- \frac{1}{12 u_s^6}{\cal R}_{ss^\prime}\right]
+
\sum_{x,s}\left[ \frac{4}{3 u_s^2  u_t^2}{\cal P}_{st} - \frac{1}{12 u_s^4 u_t^2}{\cal R}_{st}\right] \vphantom{\frac{1}{\xi}} \right\},
\end{eqnarray}
where ${\cal P}$ is the plaquette and ${\cal R}$ is the $2\times1$
rectangular Wilson loop. The coupling $g^2$ appears in $\beta=2 N_c/g^2$.
The parameter $\xi_0$ is the bare gauge anisotropy, and $u_s$ and $u_t$ are the spatial and temporal tadpole factors, dividing the spatial and temporal gauge links, respectively. This action has leading discretizations error of $O(\alpha_s^4,a_t^2,g^2 a_s^2)$ and possesses a positive definite transfer matrix since there is no length-two rectangle in time.

In the fermion sector, we adopt the anisotropic clover fermion
action~\cite{Chen:2000ej}
\begin{eqnarray}\label{eq:anisoSW}
S_F^{\xi}[U, \overline{\psi},\psi ]
&=& a_s^3 a_t \sum_{x} \overline{\psi}(x) Q \psi(x)\nonumber\\
Q &=&\left[m_0 + \nu_t {W}_t +{\nu_s} {W}_s
- \frac{a_s}{2} \left(c_{\rm t} \sigma_{st}F^{st} + \sum_{s<s^\prime} c_{\rm s} \sigma_{ss^\prime}F^{ss^\prime} \right) \right], \nonumber \\
\label{eq:orig_ferm_action}
\end{eqnarray}
where $\sigma_{\mu\nu}=\frac{1}{2}[\gamma_\mu,\gamma_\nu]$ and
\begin{eqnarray}
W_\mu &=& \nabla_\mu -\frac{a_\mu}{2} \gamma_\mu \Delta_\mu \nonumber\\
\nabla_\mu f(x) &=& \frac{1}{2a_\mu} \bigg[ U_\mu(x) f(x+\mu) -
    U^\dagger_\mu(x-\mu) f(x-\mu) \bigg] \nonumber\\
\Delta_\mu f(x) &=& \frac{1}{a_\mu^2} \bigg[ U_\mu(x) f(x+\mu) +
    U^\dagger_\mu(x-\mu) f(x-\mu)
- 2f(x) \bigg].
\end{eqnarray}
In terms of dimensionless variables $\hat \psi = a_s^{3/2}
\psi$, $\hat m_0 = m_0 a_t$, ${\hat \nabla}_\mu = a_\mu^2 \nabla_\mu$,
${\hat \Delta}_\mu = a_\mu \Delta_\mu$, $\hat F_{\mu\nu} = a_\mu a_\nu
F_{\mu\nu}$ and the dimensionless ``Wilson operator'' $\hat{W}_\mu
\equiv \hat \nabla_\mu - \frac{1}{2} \gamma_\mu \hat
\Delta_\mu$, we find the fermion matrix $Q$ becomes
\begin{eqnarray}\label{eq:pre-fermion-action}
Q & = & \frac{1}{a_t} \left\{
 a_t \hat{m_0} + \nu_t \hat{W}_t +\frac{\nu_s}{\xi_0}  \sum_s  \hat{W}_s -\frac{1}{2} \left[
      c_t \sum_{s} \sigma_{ts} \hat{F}_{ts} +
      \frac{c_s}{\xi_0}
      \sum_{s<s^\prime} \sigma_{ss^\prime} \hat{F}_{ss^\prime} \right]
      \right\}. 
\end{eqnarray}
Here $\nu$ is the ratio of the bare fermion to the bare gauge
anisotropy. From the field
redefinition~\cite{Symanzik:1983dc,Symanzik:1983gh}, there is one
redundant coefficient: either $\nu_t$ or $\nu_s$. There are two common
choices to eliminate this redundancy: setting $\nu_s = 1$ ($\nu_t$-tuning) or $\nu_t = 1$ ($\nu_s$-tuning)~\cite{Chen:2000ej}. We will use $\nu_s$-tuning in this work, so we set $\nu_t = 1$, with the
tree-level conditions on $c_s$ and $c_t$ as described in
Ref.~\cite{Chen:2000ej}. In particular, we choose the tree-level tadpole-improved values
\begin{eqnarray}
c_s = \frac{\nu}{{u}_s^3}, \quad
c_t = \frac{1}{2}\left(\nu +
\frac{1}{\xi}\right)\frac{1}{{u}_t {u}_s^2};
\label{eq:clov_coeffs}
\end{eqnarray}
where ${u}_s$ and ${u}_t$ are the tadpole factors and the fraction $a_t/a_s = 1/\xi$ is set to the desired renormalized gauge anisotropy.

In this work, the gauge links in the fermion action are 3-dimensionally
stout-link smeared gauge fields with smearing weight $\rho$ and
$n_\rho$ iterations. To distinguish tadpole factors associated with
the smeared fields appearing in the fermion action from those
appearing in the gauge action, we use notations $\tilde{u}_s$ and
$\tilde{u}_t$ for spatial and temporal tadpole factors, respectively. For convenience of parameterization, we use the bare gauge and fermion anisotropies, $\gamma_{g,f}$, defined as
\begin{eqnarray}
\gamma_g = \xi_0, \quad
\gamma_f = \frac{\xi_0}{\nu}.
\end{eqnarray}
To summarize, the final gauge and fermion actions are
\begin{eqnarray}
S_G^{\xi}[U] &=& \frac{\beta}{N_c\gamma_g} \left\{
\sum_{x, s>s^\prime} \left[  \frac{5}{3 u_s^4}{\cal P}_{ss^\prime}- \frac{1}{12 u_s^6}{\cal R}_{ss^\prime}\right]
+
\sum_{x,s}\left[ \frac{4}{3 u_s^2  u_t^2}{\cal P}_{st} - \frac{1}{12 u_s^4 u_t^2}{\cal R}_{st}\right] \vphantom{\frac{1}{\xi}} \right\}, \\
S_F^{\xi}[U, \overline{\psi},\psi ]
&=& \sum{x} \overline{\psi}(x) \frac{1}{ \tilde{u}_t} \left\{
 \tilde{u}_t \hat{m_0} +  \hat{W}_t +\frac{1}{\gamma_f}  \sum_s  \hat{W}_s -\frac{1}{2} \left[\frac{1}{2}\left(\frac{\gamma_g}{\gamma_f} + \frac{1}{\xi}
\right)\frac{1}{\tilde{u}_t \tilde{u}_s^2}
 \sum_{s} \sigma_{ts} \hat{F}_{ts} +
      \frac{1}{\gamma_f}
      \frac{1}{\tilde{u}_s^3}
      \sum_{s<s^\prime} \sigma_{ss^\prime} \hat{F}_{ss^\prime} \right]
      \right\}\psi(x). \nonumber \\
\label{eq:fermion-action}
\end{eqnarray}

\subsection{Stout-smeared links}\label{subsec:stout}

The smeared fermion action provides significant improvements on
actions that explicitly break chirality such as the clover fermion
action. It has been demonstrated that chiral symmetry is
improved~\cite{DeGrand:1998jq} after treatment of the gauge links in
the fermion action. In this work, we use three-dimensionally
stout-smeared links~\cite{Morningstar:2003gk} in the fermion
action. We use smearing parameters $\rho=0.22$ and $n_\rho=2$ (as in
Ref.~\cite{Morrin:2006tf}) through the end of
Sec.~\ref{sec:Setup}. In the numerical section, we will
examine our choice of stout-smearing parameters with greater
caution. Since the smearing does not involve the time direction, the
transfer matrix remains physical. As with other smearing techniques,
we need to check the smearing parameters carefully to avoid
potentially incorrect short-distance physics.

\begin{figure}[hbt]
\includegraphics[width=0.8\textwidth]{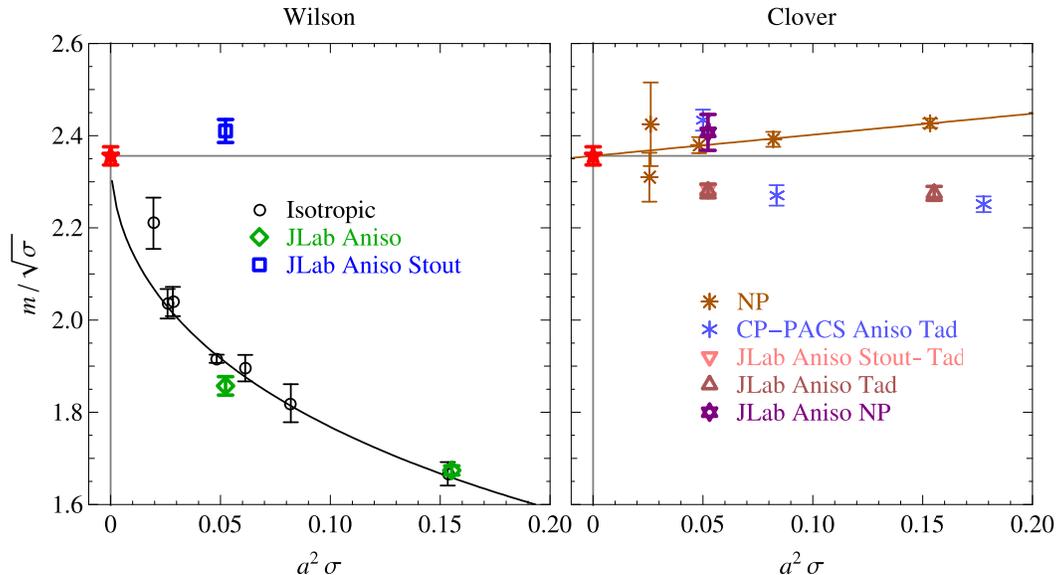}
\caption{The scaling behavior of the quenched Wilson gauge action with
Wilson (left) and clover (right) fermion actions showing the effects
of anisotropy and stout-link smearing. Shown are scaling fits of
the Wilson and clover results constrained to have the same continuum
limit~\protect\cite{Edwards:1997nh}. In the left plot, the anisotropic
stout-link smeared result is the square. In the right plot, the ``NP''
labels non-perturbatively tuned clover coefficients, and the ``Tad''
indicates tadpole-improved coefficients.
}
\label{fig:scaling}
\end{figure}

We consider the effects of our choice of action on
scaling~\cite{Edwards:1997nh} in a quenched theory. On the left-hand side of Figure~\ref{fig:scaling}, we show the scaling behavior of the vector meson mass in units of the string tension, $a m_V /\sqrt{a^2\sigma}$. In this plot, all the points on the graph have fixed quark mass determined from the ratio of the pseudoscalar to vector meson mass
ratio $m_{PS}/m_V=0.7$. The left-hand panel shows results for the quenched Wilson gauge and fermion action, and the right panel shows results for the quenched anisotropic Wilson gauge and clover fermion actions. The curves are scaling fits to the Wilson and clover fermion data constrained to have the same continuum limit~\cite{Edwards:1997nh}, with the Wilson action scaling like ${O}(a)$ and the non-perturbative clover results scaling
like ${O}(a^2)$. The horizontal line is the (fitted) continuum limit
value. The small residual scaling violations in the
nonperturbatively improved clover action (1\% at $a^2\sigma \sim 0.05$ or $a\sim0.1\mbox{ fm}$) indicate that the dominant source of
scaling violations in the Wilson action comes from chiral symmetry breaking at ${O}(a)$. Our simulations on anisotropic lattices show
similar scaling for the Wilson fermion action (the diamonds), and when we add stout-link smearing in the fermions (the squares), we see a large reduction in the scaling violations. In the clover case (on the right-hand side of Figure~\ref{fig:scaling}), similar tests were performed, with
both tadpole-improved perturbatively determined and nonperturbatively
determined clover coefficients. When stout-link smearing is added to
the anisotropic clover action, the scaling remains consistent with the unsmeared results. Thus, the three-dimensional stout-link smearing does not adversely effect the scaling properties of the fermion action, and the resulting scaling violations after smearing are suitably small.

\subsection{Schr\"odinger functional}\label{subsec:SF}

The Schr\"odinger-functional
scheme~\cite{Luscher:1992an,Luscher:1996sc,Luscher:1996ug} allows for
simulations at small pion mass, since the background field lifts zero
modes.  We use the Schr\"odinger functional and the PCAC relation to
check how close our $c_{\rm SW}$ in the fermion action is to the
nonperturbative value. In previous work with dynamical fermions, the
Alpha collaboration used two-flavors~\cite{Jansen:1998mx} with the
Wilson gauge action, and CP-PACS used two-flavor and three-flavor
calculations with the Wilson and Iwasaki gauge
actions~\cite{Umeda:2003pj,Yamada:2004ja,Aoki:2005et}. In this work,
for our calculations with the background field in the spatial direction where we have length-two rectangles in the gauge action, we used the
Schr\"odinger-functional setup from Ref.~\cite{Klassen:1997jf}.

The bare PCAC quark mass is calculated through the PCAC relation using
zero momentum projected correlators as
\begin{eqnarray}\label{eq:pcac_m}
m(x_0) & = & r(x_0) + a c_A s(x_0),
\label{eq:mass_profile}
\end{eqnarray}
where $x_0$ is some time slice (or possibly space slice) of the correlator, $c_A$ is the $O(a)$-improved coefficient for the axial current and
\begin{eqnarray}
r(x_0) & = & \frac{1}{4} \frac{\left( \partial_0 + \partial_0^* \right) f_A(x_0)}{f_P(x_0)} \nonumber \\
s(x_0) & = & \frac{1}{2} a \, \frac{\partial_0 \partial_0^* f_P(x_0)}{f_P(x_0)}.
\end{eqnarray}
The axial current $f_{A}$ with $\Gamma=\gamma_5\gamma_\mu$ (or pseudoscalar density $f_P$ with $\Gamma=\gamma_5$) is a correlation function of bulk fields ($\overline{\psi}$, $\psi$) and boundary fields at $t=0$ ($\overline{\eta}$, $\eta$) defined as
\begin{eqnarray}
f_{O_\Gamma} (t)&=& \frac{1}{V}\sum_{\bold x} \langle \overline{\psi}\Gamma\psi ({\bold x})\sum_{{\bold y},{\bold z}} \overline{\eta}({\bold y})\Gamma\eta({\bold z}) \rangle/(N_f^2-1).
\end{eqnarray}
Similarly, correlators propagating from the other wall lead to
definitions of $r^\prime$ and $s^\prime$, where the
$f_{A(P)}^\prime$ now involves the other boundary fields at $t=T$
($\overline{\eta}^\prime$, $\eta^\prime$) and a sign change.

On an isotropic lattice, one can also determine the clover
coefficient $c_{\rm SW}$ in Schr\"odinger-functional scheme from the PCAC
relation. For an arbitrary set of action parameters, the relation
$m(y_0)=m^{\prime}(y_0)$ is not satisfied in general is not satisfied for a
generic $y_0$. We define an intermediate (before nonperturbative tuning) $c_A$ as $\hat{c}_A$ according to
\begin{eqnarray}\label{eq:c_A}
\hat{c}_A(y_0)=\frac{1}{a}\frac{{r^\prime(y_0)-r(y_0)}}{{s^\prime(y_0)-s(y_0)}}
\end{eqnarray}
and define a modified mass in Eq.~\ref{eq:pcac_m} as
\begin{eqnarray} \label{eq:M_redef}
M(x_0,y_0) &=& r(x_0) - \hat{c}_A(y_0) s(x_0) \nonumber\\
M^\prime(x_0,y_0) &=& r^\prime(x_0)-\hat{c}_A(y_0) s^\prime(x_0).
\end{eqnarray}
One nonperturbatively determines $c_{\rm SW}$ by imposing the condition
\begin{eqnarray}
\Delta M = M(x_0,y_0)-M^\prime(x_0,y_0) = \Delta M^{(0)},
\label{eq:nonpt_renorm_cond}
\end{eqnarray}
where $\Delta M^{(0)}$ is either set to be zero or the tree-level mass
splitting which can be obtained from a free-field simulation with the
same setup of the gauge and fermion actions. Note that the choice of
$(x_0,y_0)$ is $(T/2,T/4)$ for both $\Delta M$ and $M$.  $c_{\rm SW}$
is obtained by tuning the condition $\Delta M^{(0)}-\Delta M=0$, and
$c_A$ is obtained in straightforward fashion from Eq.~\ref{eq:c_A}
with the correct $c_{\rm SW}$.
Note that all applications of the Schr\"odinger functional so far have
been limited to isotropic lattices.

\begin{figure}[hbt]
\includegraphics[width=0.49\textwidth]{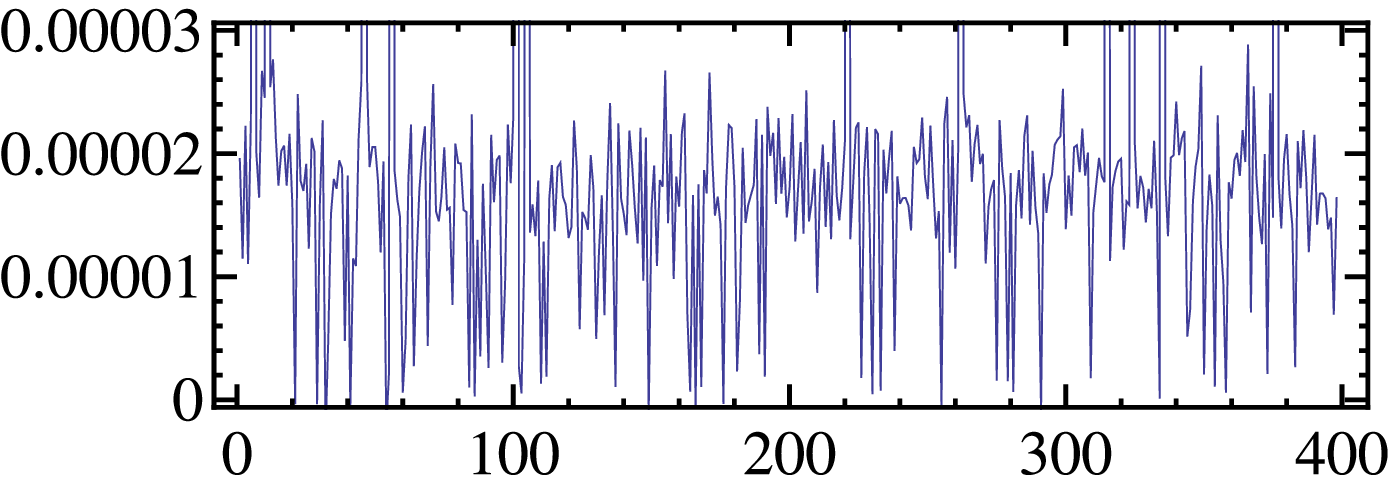}
\includegraphics[width=0.49\textwidth]{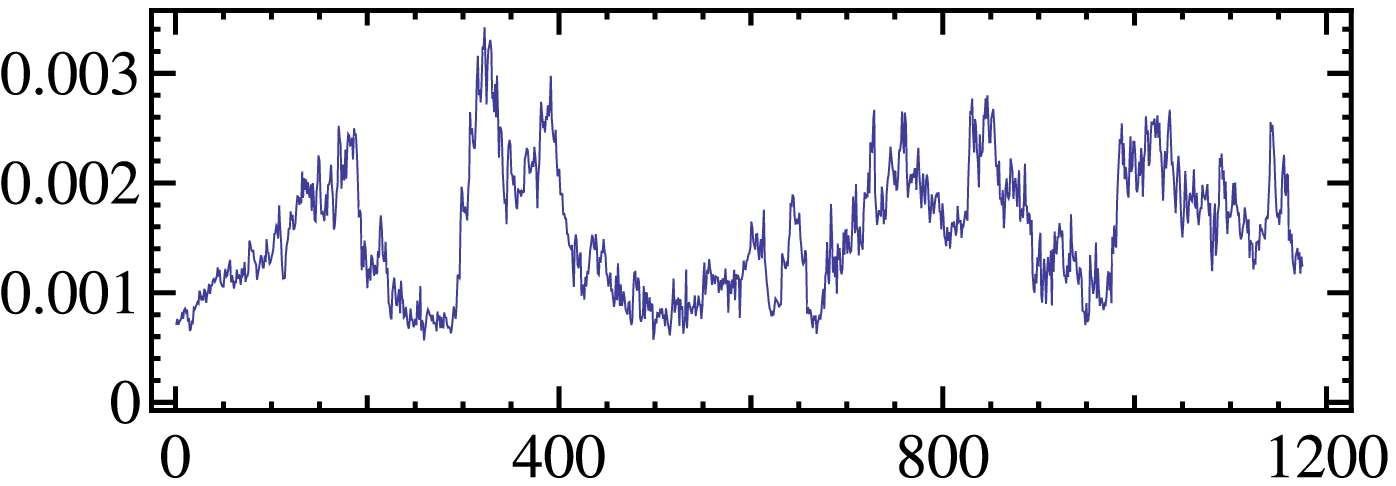}
\caption{$\lambda_{\rm min}(Q^2)$ measured in three-flavor anisotropic
clover simulations (using parameters $\beta=2.2$, $\rho=0.22$,
$n_\rho=2$, $m_0=-0.054673$, $\gamma_f=\xi=3.5$) with (left) and without
(right) a background field. The $x$-axis is in units of 5 trajectories. The right-hand panel (no background field) has a longer autocorrelation time compared to the case with a background field (left side).
The vertical scale, however, is different because of the background field.}
\label{fig:lambda}
\end{figure}

\begin{figure}[hbt]
\includegraphics[width=0.7\textwidth]{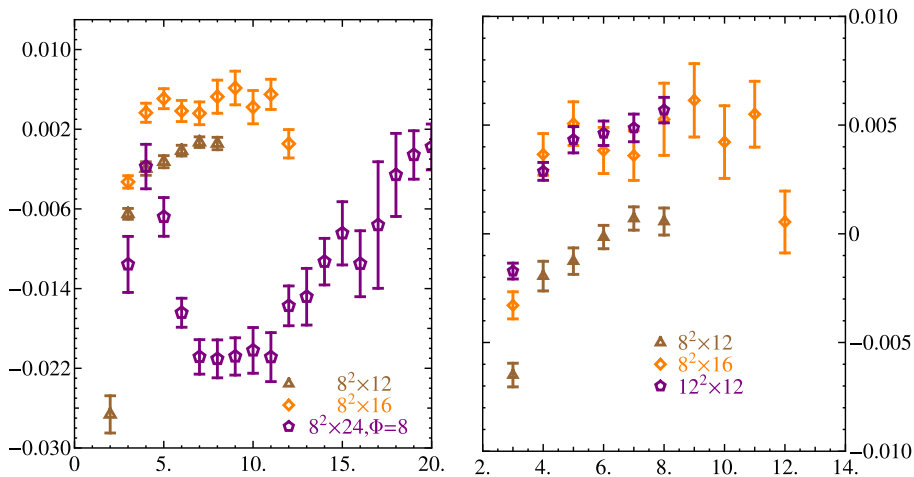}
\caption{$a_s M_s$ comparison among different spatial volumes. The
parameters in the action are the same as Figure~\ref{fig:lambda}.
The right panel is an enlarged scale of the left panel with the
addition of the $12^3$ spatial-volume result, which is the size used
in the remainder of this work.}
\label{fig:size-test}
\end{figure}
In this work, we implement the Schr\"odinger-functional setup on
{\it anisotropic} lattices for the first time in dynamical
simulations. In one of our earlier three-flavor (anisotropic) clover action simulations (with parameters $\beta=2.2$, $\rho=0.22$, $n_\rho=2$,
$m_0=-0.054673$, $\gamma_f=\xi=3.5$), we found that the autocorrelation time of the lowest eigenvalue of $Q^\dagger Q$ (as defined in Eq.~\ref{eq:pre-fermion-action}) is significantly reduced by introduction of the background field, as shown in Figure~\ref{fig:lambda}. Thus, for these background-field calculations, we can use correspondingly fewer trajectories compared to the calculations with antiperiodic boundary conditions.

We also implement the background field not only in the $t$ direction
(as conventionally used in Schr\"odinger functional) but also the $z$
direction. (We label the modified masses $M_s$ and $M_t$ with respect to the direction of the boundary field direction and similarly for the mass
difference $\Delta M_s$ and $\Delta M_t$.) We measure $M(x_0)$ with
various spatial $x$ and $y$ sizes, $L_{x,y}$, and sizes $L_z$,
showing results in Figure~\ref{fig:size-test}, all with time extent
$L_t=32$ and equal amount of statistics. When we increase the length in the $z$ direction, a good signal appears from $t=12$ to 16 but not beyond 24. This is because the background field becomes too weak at large $L_z$. When we increase the background field signal, which is proportional to $\Phi$ at $L_z=24$, the signal is still not as good as for $L_z=16$. Similar checks can be done regarding the size of $L_{x,y}$. As we increase the value from 8 to 12, the signal shows improvement. The right panel of
Figure~\ref{fig:size-test} zooms in the scale as seen from the left panel, and demonstrates that the $L_{x,y}=L_z=12$ spatial volume has the
lowest statistical error while also providing enough spatial size to adequately resolve the potential which will be discussed in Section~\ref{subsec:gauge_aniso};
this is the spatial volume we use in the remainder of this work.

We note that the fermion algorithm we will use corresponds to
simulating the $N_f=3$ fermion determinant as $\det \left( Q^\dagger Q
\right)^{\frac{3}{2}}$.  In particular, when the renormalized quark
mass is negative (below the chiral limit), there is no phase that
should arise as in a $N_f=3$ version of QCD. Thus, the algorithm is
implementing the absolute value of the fermion determinant.  While,
strictly speaking, this is not a $N_f=3$-flavor version of QCD when
the (renormalized) quark mass is negative, it does allow us to
implement a $SU(3)$ flavor symmetric version of PCAC.

\subsection{Renormalization conditions}

In principle, we can determine the critical values for the
bare parameters $\gamma_g^*$ and all the unknown
coefficients in the fermion action nonperturbatively by imposing the
following conditions:
\begin{eqnarray}\label{eq:5conditions}
\xi_g(\gamma_g^*, \gamma_f^*, m_0^*, c_s^*, c_t^*) &=& \xi \nonumber\\
\xi_f(\gamma_g^*, \gamma_f^*, m_0^*, c_s^*, c_t^*) &=& \xi \nonumber\\
M_t(\gamma_g^*, \gamma_f^*, m_0^*, c_s^*, c_t^*) &=& m_q \nonumber\\
\Delta M_s(\gamma_g^*, \gamma_f^*, m_0^*, c_s^*, c_t^*) &=& \Delta M_s^{(0)} \nonumber\\
\Delta M_t(\gamma_g^*, \gamma_f^*, m_0^*, c_s^*, c_t^*) &=& \Delta M_t^{(0)},
\end{eqnarray}
where $\xi_g$ is the renormalized gauge anisotropy, $\xi_f$ is the
renormalized fermion anisotropy (defined through the meson dispersion
relation $E(p)^2=M^2+p^2/\xi_f^2$) and $M_t$ is the PCAC quark mass
measured in Schr\"odinger-functional background field (from
Eq.~\ref{eq:pcac_m}).  $\Delta M_{s,t}$ are the mass differences
measured in two different background field directions.  However, this
requires that we search a five-dimensional parameter-space with
dynamical simulations which is costly.

In this work, we set the spatial and temporal clover coefficients
$c_s$ and $c_t$ (in Eq.~\ref{eq:fermion-action}) to the tree-level
tadpole-improved values. (Later in the numerical section, we
demonstrate using these values that the last two conditions in
Eq.~\ref{eq:5conditions} do hold within a few percent.) To determine
the remaining three coefficients, we parameterize $\xi_g$, $\xi_f$ and
$M_t$ as functions $f_i(\gamma_g, \gamma_f, m_0)$. For simplicity, we
choose functions of the bare parameters with the form
\begin{eqnarray}\label{eq:3conditions}
\xi_g(\gamma_g, \gamma_f, m_0) &=& a_0 + a_1 \gamma_g + a_2\gamma_f + a_3 m_0 \nonumber\\
\xi_f(\gamma_g, \gamma_f, m_0) &=& b_0 + b_1 \gamma_g + b_2\gamma_f + b_3 m_0 \nonumber\\
M_t(\gamma_g, \gamma_f, m_0) &=& c_0 + c_1 \gamma_g + c_2\gamma_f + c_3 m_0 .
\end{eqnarray}
These parameterizations are linear functions in the
coefficients. We can choose higher powers of the bare parameters such as a $m_0^2$ term; however, the coefficients that we need to determine would remain linear.

Once the coefficients $a_i$, $b_i$, and $c_i$ are determined,
we impose our renormalization conditions
\begin{eqnarray}\label{eq:solver}
\xi_g(\gamma_g^*, \gamma_f^*, m_0^*) &=& \xi \nonumber\\
\xi_f(\gamma_g^*, \gamma_f^*, m_0^*) &=& \xi \nonumber\\
M_t(\gamma_g^*, \gamma_f^*, m_0^*) &=& m_q
\end{eqnarray}
to obtain the critical values for the bare parameters as a function of
the input quark mass $m_q$. If only linear terms in the bare parameters are used, then the intersection of these three hyperplanes is the solution of a $3\times 3$ linear system of equations. If higher-order terms are used, then the intersection is the root of a system of functions. The fitted parameters determine the chiral limit when the input $m_q=0$.

\section{Algorithm}\label{sec:Algorithm}
Our configurations were generated with the Rational Hybrid Monte Carlo
(RHMC) algorithm~\cite{Clark:2006wq,Kennedy:2006ax,Clark:2004cq}. Strictly, RHMC refers only to the method for simulating odd flavors of fermions, and one can combine several orthogonal algorithmic improvements with the RHMC scheme resulting in a wide variety of possible RHMC
algorithms. In order to be specific therefore, we describe the RHMC
method in brief below and then detail our particular combination of
improvements.

\subsection{Rational Hybrid Monte Carlo}
The basic technique for gauge generation is a Markov Chain Monte Carlo
method, where one moves from an initial gauge configuration to a
successive one by generating a new trial configuration and then
performing an acceptance/rejection test upon it. If the trial
configuration is accepted, it becomes the successive configuration in
the chain, otherwise, the original configuration becomes the next state in the chain.

In order to use a global Metropolis accept/reject step with a
reasonable acceptance rate, the space of states is extended to include
momenta $\pi_{\mu}(x)$ canonical to the gauge links $U_{\mu}(x)$ so
that one may define a Hamiltonian
\begin{equation}
H = \frac{1}{2} \sum_{x,\mu} \pi_{\mu}(x)^\dagger \pi_{\mu}(x) + S(U)
\end{equation}
where $S$ is the action. It is then possible to propose new
configurations from previous ones by performing Molecular Dynamics
(MD). Using a reversible and area preserving MD step maintains
detailed balance, which is sufficient for the algorithm to
converge. In order to ensure ergodicity in the entire phase space, the
momenta need to change periodically. This can be affected by
refreshing the momenta from a Gaussian heatbath prior to the MD update
step.

In order to deal with the fermion determinant, it is standard to use
the method of pseudofermions. One integrates out the Grassman-valued
fermion fields in the action and rewrites the resulting determinant as
an integral over bosonic fields
\begin{equation}\label{eq:FermPartition}
Z = \int [d\bar{\eta}] [d\eta] e^{-\bar{\eta} \mathcal{D} \eta} =
\det\left( \mathcal{D} \right) = \int [d\phi^\dagger] [d \phi]
e^{-\phi^\dagger \mathcal{D}^{-1} \phi}
\end{equation}
where $\eta$ and $\bar{\eta}$ are the Grassman valued fields,
$\mathcal{D}$ is some fermionic kernel and $\phi^\dagger$ and $\phi$
are the bosonic pseudofermion fields. Our phase space is thus enlarged
to include also the pseudofermion fields, which similarly to the
momenta, need to be refreshed before each MD step, to carry out the
the pseudofermion integral.

In the case of a two-flavor simulation, $\mathcal{D}$ is typically of
the form
\begin{equation}
\mathcal{D} = Q^\dagger Q
\end{equation}
where $Q$ is the fermion matrix for an individual flavor of
fermion. In this case $\mathcal{D}$ is manifestly Hermitian and
positive definite, and the integral in Eq.~\ref{eq:FermPartition} is
guaranteed to exist. Furthermore, the pseudofermion fields can easily
be refreshed by producing a vector $\chi$ filled with Gaussian noise
with a variance of $\frac{1}{2}$ and then forming $\phi = Q^\dagger
\chi$.

In the case of an odd number of flavors, since $Q$ itself is not
guaranteed to be positive definite, one works instead with
$\sqrt{\left(Q^\dagger Q\right)}$:
\begin{equation}
\det\left( Q \right) = \det \left( Q^\dagger Q \right)^{\frac{1}{2}} =
\int [d\phi^\dagger][d \phi] e^{-\phi^\dagger \left(Q^\dagger
Q\right)^{-\frac{1}{2}} \phi} \ .
\end{equation}
The square root in $\mathcal{D}$ can be approximated to numerical
precision using a low-order rational approximation
$r^{-\frac{1}{2}}(Q^\dagger Q)$ which can be expressed in a partial fraction (sum over poles) form as
\begin{equation}
\left(Q^\dagger Q\right)^{-\frac{1}{2}} \approx r^{-\frac{1}{2}}(Q^\dagger Q) =
\alpha_0 I + \sum_{k} \alpha_k \left[ Q^\dagger Q + \beta_k
\right]^{-1} \ ,
\end{equation}
where the rational function approximation is specified by the
coefficients $\alpha_k$ and $\beta_k$. In particular the action of all the
denominator pieces onto one vector $\phi$ involves the solution of
linear systems
\begin{equation}
 \left[ Q^\dagger Q + \beta_k \right] \chi = \phi
\end{equation}
which can be performed simultaneously using a multiple-shift (a.k.a.
multi-mass) conjugate-gradient solver~\cite{Jegerlehner:1996pm}. Refreshment of the $\phi$ field now
proceeds by evaluating
\begin{eqnarray}
\phi = r^{\frac{1}{4}}(Q^\dagger Q) \eta
\end{eqnarray}
where the $\eta$ are once again filled with Gaussian noise of variance
$\frac{1}{2}$, and $r^{\frac{1}{4}}$ is now a rational approximation to
$\left( Q^\dagger Q \right)^{\frac{1}{4}}$.  Molecular dynamics forces
now need to be calculated for each pole term in the partial fraction
\begin{equation}
F_{\rm rational} = -\sum_k \alpha_k \ \phi^\dagger \left(Q^\dagger Q
+\beta_k \right)^{-1} \left[ \frac{d Q^\dagger}{dU} Q + Q^\dagger
\frac{dQ}{dU} \right] \left( Q^\dagger Q +\beta_k \right)^{-1} \phi .
\end{equation}

The idea of using a rational approximation in the action, and its
consequences for field refreshment and molecular dynamics forces make
up the basics of the RHMC algorithm.

In terms of tuning, in our single-precision simulations we tune the
rational approximation coefficients by requiring that the
approximation has a maximum error over the approximation interval that
is smaller than the solver residua we require in the evaluation of the
partial fraction expansions, namely $10^{-8}$ for energy calculations
and $10^{-6}$ for the force calculations. Further, in the case of the
force calculations we successively relax the solver criteria for poles
that have smaller contributions to the force as in
Refs.~\cite{Clark:2006wq,Kennedy:2006ax,Clark:2004cq}.

\subsection{Multi-Scale Anisotropic Molecular Dynamics Update}
While any reversible and area-preserving MD update scheme can be used
in the MD step, the acceptance rate is controlled by the truncation
error in the scheme. This manifests itself as a change in the
Hamiltonian, $\delta H$, over an MD trajectory, since we use the
Metropolis acceptance probability
\begin{equation}
P_{\rm acc} = {\rm min}\left( 1, e^{-\delta H} \right) \ .
\end{equation}
We may easily construct a manifestly reversible scheme by combining
symplectic update steps $\mathcal{U}_{p}(\delta \tau)$ and
$\mathcal{U}_{q}(\delta \tau)$ which update momenta and coordinates by
a time step of length $\delta \tau$ respectively
\begin{eqnarray}
  \mathcal{U}_{p}(\delta \tau_{\mu}) : \ \left( \pi_{\mu}(x),
  U_{\mu}(x) \right) & \rightarrow & \left( \pi_{\mu}(x) +
  F_{\mu}(x)\delta \tau_{\mu}, U_{\mu}(x) \right) \\
  \mathcal{U}_{q}(\delta \tau_{\mu}) : \ \left( \pi_{\mu}(x),
  U_{\mu}(x) \right) & \rightarrow & \left( \pi_{\mu}(x) , e^{i \pi
  \delta \tau_{\mu}} U_{\mu}(x) \right),
\end{eqnarray}
where $F_{\mu}(x)$ is the MD force coming from the variation of the action with respect to the gauge fields. We emphasize that one may update all the links pointing in direction $\mu$ with a separate step-size $\delta \tau_{\mu}$. While this may not be useful in isotropic simulations, in an anisotropic calculation with one fine direction, it may be advantageous to use a shorter timestep to update the links in that direction to ameliorate the typically larger forces that result from the shorter lattice spacing~\cite{Morrin:2006tf}. The anisotropy in step size requires a small amount of manual fine tuning, but should be similar to the
anisotropy in the lattice spacings.

Our base integration scheme in this work is due to
Omelyan~\cite{Omelyan:2003,deForcrand:1996ck,Sexton:1992nu}; we
use the combined update operator
\begin{equation}
\mathcal{U}^1(\delta \tau) = \mathcal{U}_p(\lambda \delta \tau)
\mathcal{U}_q( \frac{1}{2} \delta \tau ) \mathcal{U}_{p}(1 - {2}
\lambda \delta \tau) \mathcal{U}_{q}( \frac{1}{2} \delta \tau )
\mathcal{U}_p( \lambda \delta \tau)
\end{equation}
which results in a manifestly reversible scheme that is accurate to
$O(\delta \tau^3)$. The size of the leading error term can be further
minimized by tuning the parameter $\lambda$. In our work we used the
value of $\lambda$ from Ref.~\cite{deForcrand:1996ck} without any further
tuning, which promises an efficiency increase of approximately 50\%
over the simple leapfrog algorithm.

In Refs.~\cite{Sexton:1992nu,Weingarten:1980hx} it was shown that a
reversible, multi-level integration scheme can be constructed which
allows various pieces of the Hamiltonian to be integrated at different
timescales. Let us consider a Hamiltonian of the form
\begin{equation}
H(\pi, U)=\frac{1}{2} \pi^\dagger_{\mu}(x) \pi_{\mu}(x) + S_1(U) + S_2(U)
\end{equation}
where $S_1(U)$ and $S_2(U)$ are pieces of the action with corresponding
MD forces $F_1$ and $F_2$ respectively. One can then split the
integration into 2 timescales. One can integrate with respect to
action $S_1(U)$ using $\mathcal{U}^1(\delta \tau_1)$, where in the
component $\mathcal{U}_{p}(\delta \tau_1)$ we use only the force $F_{1}$. The
whole system can then be integrated with the update
\begin{equation}
\mathcal{U}^2(\delta \tau_2) = \mathcal{U}^\prime_p(\lambda \delta \tau_2)
\mathcal{U}^1( \frac{1}{2} \delta \tau_2 ) \mathcal{U}^\prime_{p}(1 - {2}
\lambda \delta \tau_2) \mathcal{U}^1( \frac{1}{2} \delta \tau_2 )
\mathcal{U}^\prime_p( \lambda \delta \tau_2)
\end{equation}
where in $\mathcal{U}^\prime_p$ we update the momenta using only $F_2$. Thus we end up with two characteristic integration timescales $\delta
\tau_1$ and $\delta \tau_2$. The scheme generalizes recursively to a
larger number of scales. A criterion for tuning the algorithm is to
arrange for terms in the action to be mapped to different timescales
so that on two timescales $i$ and $j$ we have $|| F_i || \delta
\tau_{i} \approx ||F_j|| \delta \tau_j$, as suggested in
Ref.~\cite{Hasenbusch:2002ai}.  We now proceed to outline how we split
our action.

\subsection{Even-Odd Preconditioning the Clover Term}
The clover term may be preconditioned by labeling sites as even and
odd, and grouping together the terms in the operator that connect
sites of various labels. In particular
\begin{equation}
Q = \left( \begin{array}{cc}
A_{ee} & D_{eo} \\
D_{oe} & A_{oo} \end{array} \right),
\end{equation}
where the blocks $A_{ee}$ and $A_{oo}$ contain the clover term and the
diagonal parts of the Wilson operator whereas the $D_{eo}$ and
$D_{oe}$ contain the Wilson hopping term. This matrix can be block
diagonalized as
\begin{equation}
Q =\left( \begin{array}{cc}
  1 & 0 \\
D_{oe} A^{-1}_{ee} & 1
\end{array}
\right)
\left( \begin{array}{cc}
A_{ee} & 0 \\
0 & A_{oo} - D_{oe} A^{-1}_{ee} D_{eo} \\
\end{array} \right)
\left( \begin{array}{cc}
1 & A^{-1}_{ee} D_{eo} \\
0 & 1 \\
\end{array} \right),
\end{equation}
and it is clear that
\begin{equation}
\det \left(Q\right) = \det \left( A_{ee} \right) \det \left( A_{oo} -
D_{oe} A^{-1}_{ee} D_{eo} \right).
\end{equation}
We can write an action containing $N_f$ degenerate fermion flavors as
\begin{equation}
\det \left( Q^\dagger Q \right)^{\frac{N_f}{2}}
= e^{\sum_{x} N_f \tr \log A_{ee}(x)
- \sum_{i=1}^{N_f} \phi^\dagger_i \left[ r^{-\frac{1}{2}}(\tilde{Q}^\dagger \tilde{Q}) \right] \phi_i }
\end{equation}
with the preconditioned fermion matrix
\begin{equation}
\tilde{Q} = A_{oo} - D_{oe} A^{-1}_{ee} D_{eo}.
\end{equation}

This manner of preconditioning the clover term is quite standard; guides to implementation are detailed in Ref.~\cite{Hasenbusch:2002ai}. We mention that it may be possible to combine the three pseudofermion terms each containing $r^{-\frac{1}{2}}(\tilde{Q}^\dagger \tilde{Q})$ into a single one containing instead a single rational approximation $r^{-\frac{3}{2}}$, however the Remez algorithm for this approximations results in negative roots with dire implications for our multi-shift solver. In this work therefore, we have simulated with three separate one-flavor pseudofermion terms.

It is our experience that the forces resulting from the $\tr \log A_{ee}$ terms were at least an order of magnitude smaller than the
pseudofermion terms, and so the two kind of terms could be run on
separate timescales.

\subsection{Stout-Link Smearing in Fermion Actions}
The fermionic terms in our action employ stout-link smearing~\cite{Morningstar:2003gk} on the links in the spatial direction. We leave the temporal direction unsmeared to keep the transfer matrix physical. The Schr\"odinger-functional boundary condition is imposed at every iteration of the stout-link smearing. Our fermion operator $\tilde{Q}$ is evaluated on the stout-smeared fields. In the calculation of the fermion forces, we compute the force on the stouted links, but then have to also apply the chain rule to compute the force coming from the original thin links
\begin{equation}
\frac{d \tilde{Q}}{d U_{\rm thin} } = \frac{ d\tilde{Q} }{d U_{\rm stout}}
\frac{ d U_{\rm stout} }{ d U_{\rm thin}}
\end{equation}
In particular the $\frac{ d U_{\rm stout}}{ d U_{\rm thin} }$ term is
common to all the poles in the rational function approximation; hence,
in our rational forces computations we compute all the forces with
respect to the stout links first and then perform the recursion to the
thin links only once. The Schr\"odinger-functional boundary conditions
imply that the force for the links which we hold fixed are set to
zero.

\subsection{Split Gauge Term}
We can write our gauge action schematically as
\begin{equation}
S = S_s(U) + S_t(U),
\end{equation}
where the term $S_s$ contains only loops with spatial gauge links, and
the $S_t$ term contains loops contain spatial and temporal
links. While the term $S_s$ produces forces only in the spatial
directions, the $S_t$ term produces forces in both the spatial and the
temporal directions. In particular the spatial forces from $S_t$ are
larger in magnitude than the spatial forces from $S_s$ by roughly the
order of the anisotropy, and in turn, the temporal forces from $S_t$
are larger than the spatial forces from $S_t$. Our anisotropic
integration step-size balances the spatial and temporal forces of the
$S_t$ term against each other. However, in order to balance the spatial
forces from $S_t$ and $S_s$ against each other, we integrate them on
separate time scales.

\subsection{Summary}
To summarize, we use the RHMC algorithm, with approximations tuned
separately for the force and energy calculations. Our molecular
dynamics scheme uses anisotropic timesteps and a recursively defined
multi-level integration scheme based on Omelyan's inverter. Our
fermionic terms employ stout smearing in the spatial directions only.
We have 4 kinds of terms in our molecular dynamics integration: 
the $\tr \log A_{ee}$ term, the $\phi^\dagger_i r^{-\frac{1}{2}}(\tilde{Q}) \phi$ pseudofermionic terms, the spatial
gauge action term $S_s$ and the temporal gauge action term $S_t$.

Based upon the magnitude of the molecular dynamics forces, we split
our integration scheme onto three time scales
\begin{itemize}
\item
Time scale 1 uses the Omelyan integrator with a timestep of $\delta t_1$, and
contains the $\tr \log A_{ee}$ term and the pseudofermion terms.
\item
Time scale 2 uses a leapfrog integrator with a timestep of $\delta t_2$ relative
to time scale 1, and contains the spatial gauge action term.
\item
Time scale 3 uses a leapfrog integrator with a timestep of $\delta t_3$ relative
to time scale 2, and contains the temporal gauge action term.
\end{itemize}
For our $12^3\times 32$ volume results with a background field, we used timesteps
$(\delta t_1,\delta t_2, \delta t_3) = (1/4,1/4,1/3)$ for the three time scales,
and for our $12^3\times 96$
volume results with antiperiodic boundary conditions we used $(1/5,1/3,1/2)$.
In addition, the time step for the temporal direction was a factor of
$\xi=3.5$ times smaller than the spatial timesteps.
The acceptance rate was typically between 60 and 70\%.
These technologies are all implemented in the Chroma software system~\cite{Edwards:2004sx}.

\section{Numerical Results}\label{sec:Num}

We are interested in a spatial lattice spacing on the order of
0.1--0.2~fm with a target anisotropy of $\xi = 3.5$, which would
provide a fine-enough temporal lattice spacing for excited-state
physics. We proceeded by making an initial guess for the anisotropy
parameters, $\gamma_g=\gamma_f=\xi=3.5$ and $m_0=-0.05$, and computed the Sommer scale $r_0/a_s$ (Ref.~\cite{Sommer:1993ce}) over a range of $\beta$ with lattice sizes of $12^3\times32$. Anticipating significant running in $r_0/a_s$, we chose to use $\beta=1.5$. A preliminary investigation indicates a lattice spacing of roughly $0.12$~fm. However, a careful determination of the lattice spacing involves a study of the
static quark potential as well as hadron masses on $N_f=2+1$ ensembles, a
determination of the strange quark mass, extrapolation of the light
quark mass to the physical limit, and examining different approaches
to include potential systematic uncertainties which will be presented in Ref.\cite{Edwards:aniso_tune}. Note that in the earlier part
of this section, we will vary $\beta$ to make a global search and to
get an understanding of the stout-smearing parameters and tadpole
factors.

We tune the gauge and fermion anisotropies in three-flavor
simulations. The determination of the tadpole factors used in the gauge
and fermion action and the stout-link parameter study will be
described in Sec.~\ref{subsec:tadpole-Stout}. We ultimately decide
to 3d stout-link smear with $n_\rho = 2$ and $\rho =
0.14$. Sec.~\ref{subsec:gauge_aniso} and \ref{subsec:ferm_aniso} describe how the gauge and fermion anisotropies are measured, and finally the target coefficients in the fermion action are determined in Sec.~\ref{subsec:tuning}.

\subsection{Plaquette, Tadpole Factors and Stout Links}\label{subsec:tadpole-Stout}

In this work, we tadpole improve~\cite{Lepage:1992xa} the gauge
and fermion actions. This procedure amounts to replacing the gauge link
fields with $\hat{U}_\mu(x)= U_\mu(x) / u$. We obtain the tadpole factors used in the simulation from a nonperturbative tuning. There are two types of tadpole factors; we denote using $\tilde{u}_{s,t}$ the tadpole factors with the stout-smeared fields (which are used in the gauge links of the fermion action), and ${u}_{s,t}$ are those without smearing (which are in the gauge action). For each parameter set, we start with a perturbative guess, $u_s=u_t=1$, and calculate the actual $u_{s,t}$ from the
plaquette. Then, we use the new value to feed back into the next
dynamical run. The value of $u_{s,t}$ will converge, giving the
nonperturbative value with 1--2\% precision.

We calculated these tadpole factors with a large range of $\beta$
values in three-flavor simulations. Over a wide range of parameters,
we find that the square root of the spatial plaquette is consistent
with the temporal plaquette, for both smeared and unsmeared links,
indicating that the temporal tadpole factor is close to one. We chose
to fix $u_t=\tilde{u}_t=1$ throughout the remainder of the work.
It still remains to determine the spatial tadpole factors.

\begin{figure}
\includegraphics[width=0.5\textwidth]{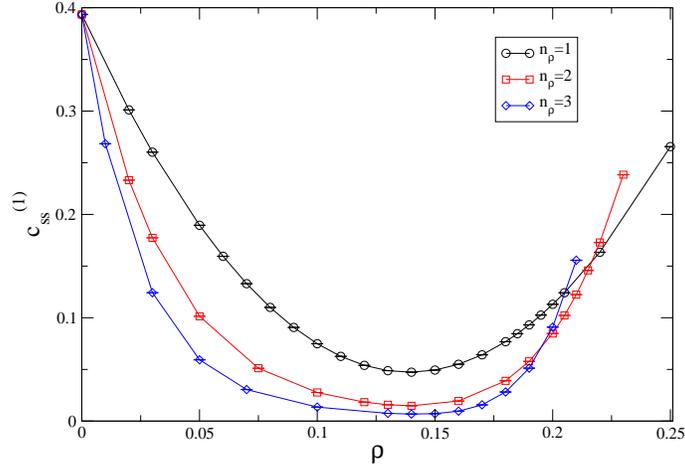}\\
\caption{The one-loop coefficient of the spatial plaquette,
$c_{ss}^{(1)}$, as a function of $\rho$ for $n_\rho=1,2,3$. The
minimum $c_{ss}^{(1)}$ (where the plaquette is near 1) in all three
cases, as shown above, indicates $\rho \approx 0.14$.}
\label{fig:spatial_plaquettes}
\end{figure}

\begin{figure}
\includegraphics[bb=42 54 716 580,width=0.5\textwidth]{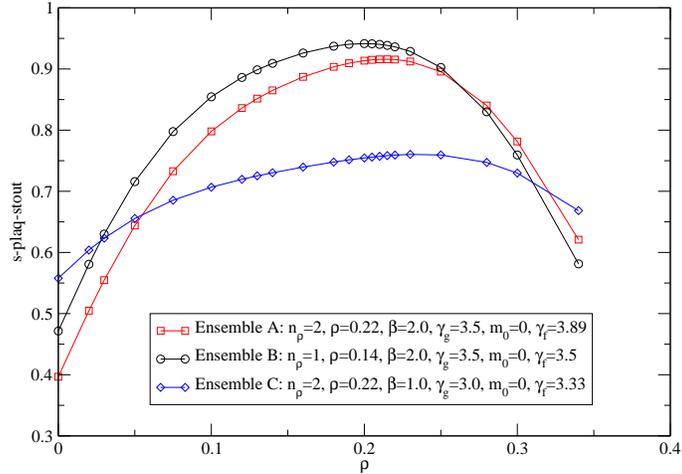}\\
\caption{This graph shows how the spatial stout-smeared plaquette
varies as a function of valence stout parameter, $\rho$, on three
different ensembles with different choices of fermion parameters,
$\beta$ and stout-smearing factors, as indicated in the legend
box. The smeared threshold is only observed for $\rho_{\rm thresh} \leq 0.2$ in $\beta=2.0$ case; for smaller $\beta$, the maximum $\rho_{\rm thresh} \approx 0.25$.
}
\label{fig:s_plaq_stout.scan}
\end{figure}

Let us first concentrate on the influence of the stout-smearing
parameters on the plaquette. In Sec.~\ref{sec:Setup}, we set the
stout-link parameters to be $\rho=0.22$ and $n_\rho=2$, following the
choice of the two-flavor anisotropic study done in Ref.~\cite{Morrin:2006tf}. At one-loop level we expect the spatial
plaquette to be
\begin{eqnarray}\label{eq:plaq}
\langle {\cal P}_s \rangle  &=& 1-c_{ss}^{(1)}g^2
\label{eq:1loop_spat_plaq}
\end{eqnarray}
where $c_{ss}^{(1)}$ is a function of the gauge action and stout-link
parameters. Figure~\ref{fig:spatial_plaquettes} shows $c_{ss}^{(1)}$
as a function of $\rho$ for $n_\rho=1,2,3$, as calculated in Ref.~\cite{Foley} with our choice of a Symanzik-improved gauge action. We find that the choice of $\rho=0.14$ corresponds to a minimum of $c_{ss}^{(1)}$ which corresponds to the maximum of the spatial plaquette in Eq.~\ref{eq:1loop_spat_plaq} for $g^2>0$. We note that this value of $\rho=0.14$ is also consistent with the suggestion~\cite{Capitani:2006ni} that the maximum allowable value, given by a classical argument, is $1/6$ in our case of three-dimensional smearing. We also investigate the spatial plaquettes in our numerical studies; see Figure~\ref{fig:s_plaq_stout.scan}. In this investigation, we use
three different three-flavor ensembles with sea-sector parameters as follow:
\begin{itemize}
\item Ensemble A: $\beta=2.0$, $\gamma_g=3.5$, $\gamma_f=3.89$, $\{\rho,n_\rho\}=\{0.22,2\}$
\item Ensemble B: $\beta=2.0$, $\gamma_g=3.5$, $\gamma_f=3.5$, $\{\rho,n_\rho\}=\{0.14,1\}$
\item Ensemble C: $\beta=1.0$, $\gamma_g=3.0$, $\gamma_f=3.33$, $\{\rho,n_\rho\}=\{0.22,2\}$
\end{itemize}
with fixed $m_0=0$, $12^3\times32$ volume with antiperiodic
boundary conditions in time. We vary $\rho$ in the valence sector
and study the resulting behavior of the spatial plaquettes. We find
that plaquette is maximized in the vicinity close to our initial
choice $\rho=0.22$.

\begin{figure}
\includegraphics[angle=-90,width=0.5\textwidth]{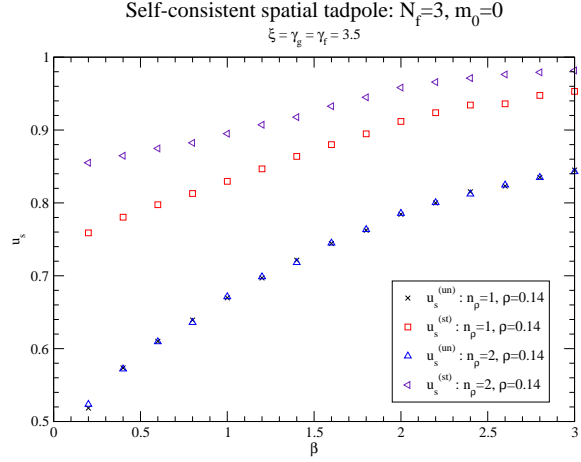}
\caption{The self-consistent tadpole factors as a functions of
$\beta$ at fixed $m_0$, $\gamma_g=\gamma_f=\xi=3.5$ and
$u_t=\tilde{u}_t=1$. The unsmeared spatial tadpole
factors $u_s^{\rm (un)}$ are nearly identical for both $n_\rho=1$ and $n_\rho=2$.}
\label{fig:tadpole_scan}
\end{figure}

We then concentrate on one set of fermion coefficients, $\rho=0.14$,
$m_0=0$, $\gamma_g=3.5$, and $\gamma_f=3.5$, and study the tadpole
factors as functions of $\beta$ with two values of $n_{\rho}$, as
shown in Figure~\ref{fig:tadpole_scan}. We see that the fermion
stout-link smeared tadpole factor $\tilde{u}_s$ (denoted as $u_s^{\rm
(st)}$ in the plot) increases when we double the number of smearing
steps and gets closer to one, which is what we expected. The gauge
tadpole factor (without stout-link smearing, denoted as $u_s^{(\rm
un)}$) remains unchanged, as if the stout-smearing in fermion sector
had no impact in the gauge sector. This tells us the ultraviolet
observable does not change with $n_{\rho}=2$; therefore, in the
rest of this paper, we will fix the smearing parameters to $\rho=0.14$
and $n_{\rho}=2$.

\begin{figure}
\includegraphics[width=0.45\textwidth]{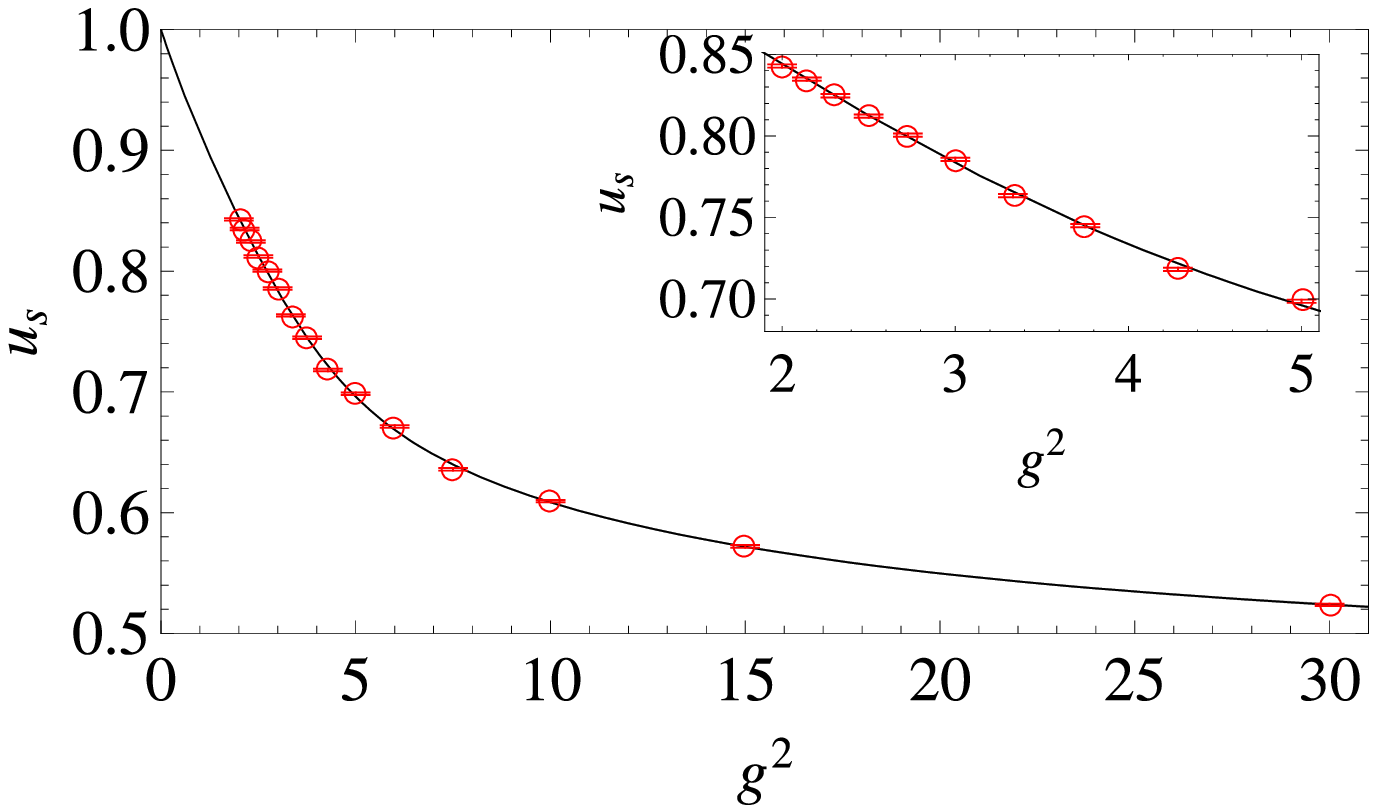}
\includegraphics[width=0.44\textwidth]{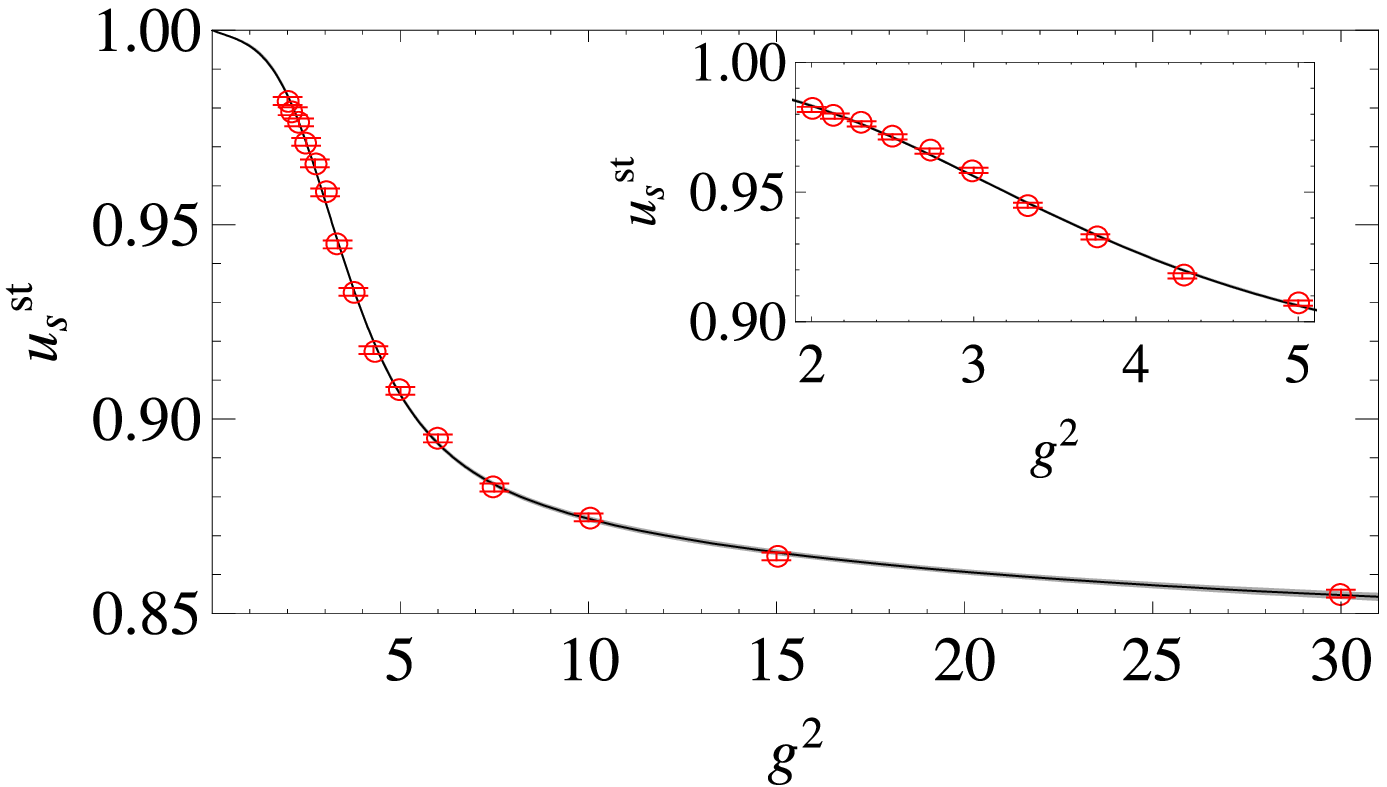}
\caption{Pad\'e approximation for tadpole factors $u_s$ and $u_s^{\rm (st)}$ for $n_\rho=2$. }
\label{fig:tadpole_pade}
\end{figure}

We parameterize the smeared and unsmeared spatial tadpole factors as
\begin{eqnarray}\label{eq:tadpole}
u &=& \sum_{n=1}^{3}\frac{1+a_n g^{2n}}{1+b_n g^{2n}}
\label{eq:Pade}
\end{eqnarray}
with the constraint $a_1-b_1=c^{(1)}$, where $g^2$ is defined as
$6/\beta$ and $c^{(1)}$ is the one-loop perturbative value of the tadpole
factor~\cite{Foley}. Figure~\ref{fig:tadpole_pade} shows the data for
$n_\rho=2$ and $\rho=0.14$ and the fit using Eq.~\ref{eq:tadpole}. The
tadpole factors are interpolated well over a large range of
$\beta$. At $\beta=1.5$ (on which we will focus for the rest of
the paper) this parameterization gives:
\begin{equation}
u_s = 0.7336, \ \qquad
\tilde{u}_s = 0.9267.
\end{equation}
In all further investigations, we use the values as determined
by Eq.~\ref{eq:Pade}. We note that as we vary $\beta$, anisotropies and
masses, the values predicted by Eq.~\ref{eq:Pade} agree
to a few percent.

\subsection{Gauge Anisotropy} \label{subsec:gauge_aniso}

We determine the gauge anisotropy from the static-quark potential using
Klassen's ``Wilson-loop ratio'' approach~\cite{Klassen:1998ua}. In this
method, we use Wilson loops involving the temporal direction, $W_{st}$,
and those in the spatial directions, $W_{ss}$, with Schr\"odinger-functional boundary
conditions applied in the $z$ direction. We measure the ratios
\begin{eqnarray}\label{eq:WL_ratio}
R_{ss}(x,y) &=&  \frac{W_{ss}(x,y)}{ W_{ss}(x+1,y)}
\xrightarrow{\rm asym} e^{-a_s V_s (y a_s)} ,
  \nonumber \\
R_{st}(x,t) &=&  \frac{W_{st}(x,t)}{W_{st}(x+1,t)}
\xrightarrow{\rm asym} e^{-a_s V_s (t a_t)} \nonumber\\
\end{eqnarray}
on lattice volumes of $12^3\times 32$. Naturally,
one should impose $R_{ss}(x,y)=R_{st}(x,t)$ to get the renormalized
$\xi_g$. An advantage to this method is that finite-volume artifacts
tend to cancel in the ratios as demonstrated in the case of the (quenched) Wilson gauge action~\cite{Klassen:1998ua} and the Iwasaki gauge action~\cite{Umeda:2003pj} with dynamical fermions. We determine $\xi_g$ by minimizing~\cite{Umeda:2003pj}
\begin{equation}\label{eq:xiR}
L(\xi_g) = \sum_{x,y}\frac{(R_{ss}(x,y)-R_{st}(x,\xi_g y))^2}
{(\Delta R_s)^2+(\Delta R_t)^2},
\end{equation}
where $\Delta R_s$ and $\Delta R_t$ are the statistical errors of $R_{ss}$ and $R_{st}$. We interpolate $R_{st}(x,t)$ by a cubic spline in
terms of $t$. To avoid short-range lattice artifacts, $x$ and $y$
should not be too small. It has been observed in the $N_f=2$
case~\cite{Umeda:2003pj} that including data with $y=1$ introduces
significant artifacts due to excited-state contamination.

\begin{figure}[htb]
\includegraphics[width=0.45\textwidth]{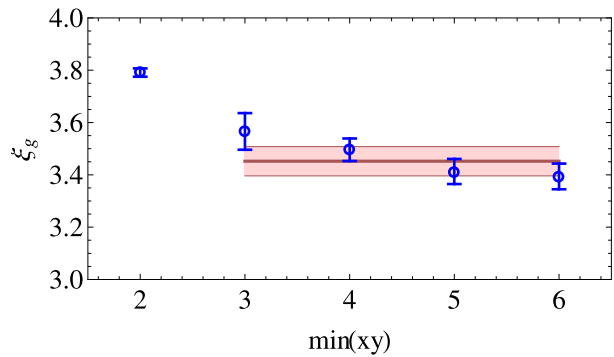}
\caption{Consistency check on $\xi_g$ measured in Schr\"odinger-functional boundary conditions (band) and periodic boundary conditions (points) as functions of the product of $x$ and $y$. The fermion action parameters for this run are $\gamma_g=4.4$, $m_0=-0.0570$ and $\gamma_f=3.3$.
}\label{fig:xig_check}
\end{figure}

While this method was originally proposed for use in gauge actions
with periodic boundary conditions, we note that it can also be applied to
the case of a constant background field. The basic observation is that
if we use Wilson loops orthogonal to the background field direction
(the boundary condition direction), the flux from the background field
that propagates should cancel in the ratios $R_{ss}$ or $R_{st}$. To
confirm this, we measured $\xi_g$ with two different three-flavor ensembles
with the same action parameters $\beta=1.5$, $\gamma_g=4.4$,
$m_0=-0.0570$ and $\gamma_f=3.3$ but under two boundary conditions:
Schr\"odinger-functional and periodic boundary conditions in the $z$
direction. Figure~\ref{fig:xig_check} shows the results from both
measurements. The gauge anisotropy from periodic boundaries,
$\xi_g^{\rm PBC}$, as a function of $\min(xy)$ is plotted. Compared
with two-flavor simulation~\cite{Umeda:2003pj}, the three-flavor
results show larger dependence on $\min(xy)$. The band in
Figure~\ref{fig:xig_check} is the resulting gauge anisotropy from
Schr\"odinger-functional boundaries, $\xi_g^{\rm SF}$, (obtained from
$\min(xy)=4$); it is consistent with $\xi_g^{\rm PBC}$. We gain
advantages by calculating $\xi_g^{\rm SF}$ instead of $\xi_g^{\rm
PBC}$ for reasons described in Sec.~\ref{subsec:SF}. Therefore, for
the rest of this work, we will use the gauge anisotropy $\xi_g^{\rm SF}$
as $\xi_g$.

\subsection{Fermion Anisotropy} \label{subsec:ferm_aniso}

We determine the fermion anisotropy $\xi_f$ through the conventional
relativistic meson dispersion relation
\begin{equation}
E(\vec{p})^2 = m^2 + \frac{\vec{p}^2}{\xi_f^2}
\label{eq:disp}
\end{equation}
where the energy $E$ and the mass $m$ are in units of $a_t$, and
$\vec{p}=2\pi\vec{n}/L_s$, with $L_s$ the spatial lattice size, is in
units of $a_s$.  Gauge configurations are generated using periodic
boundary conditions in space and antiperiodic in time on lattice volumes
of $12^3\times 96$. We use six equally separated time sources,
two different source quark smearings and an unsmeared (local) sink to
produce the correlators. These are then averaged over the six
time sources. The two resulting hadron correlators for each momentum,
averaged over equivalent rotations, were used in a constrained fit
to two amplitudes and one mass. These results are also cross-checked
against a fit including two masses (a ground and excited state).

\begin{figure}[htb]
\includegraphics[width=0.45\textwidth]{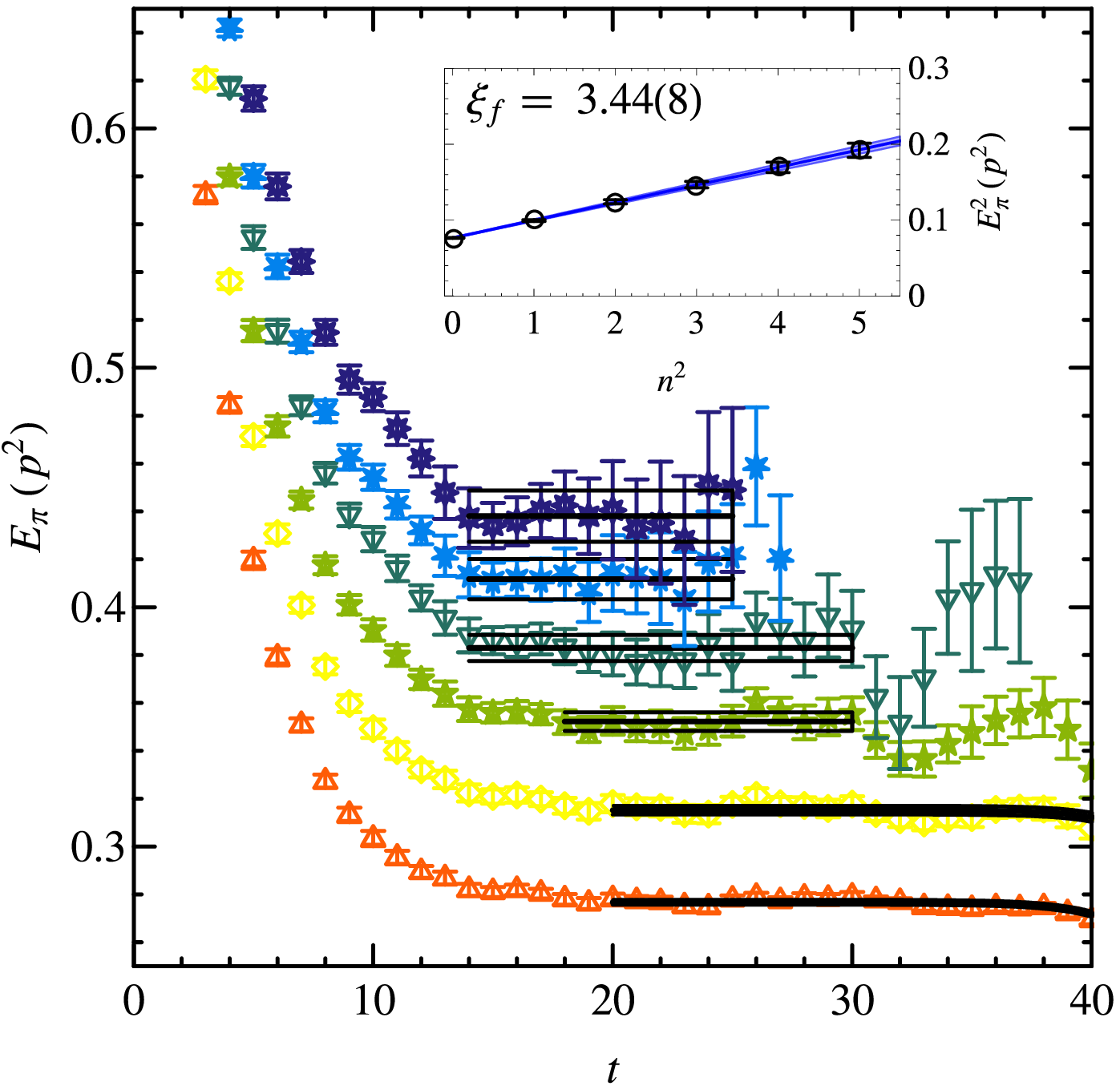}
\includegraphics[width=0.45\textwidth]{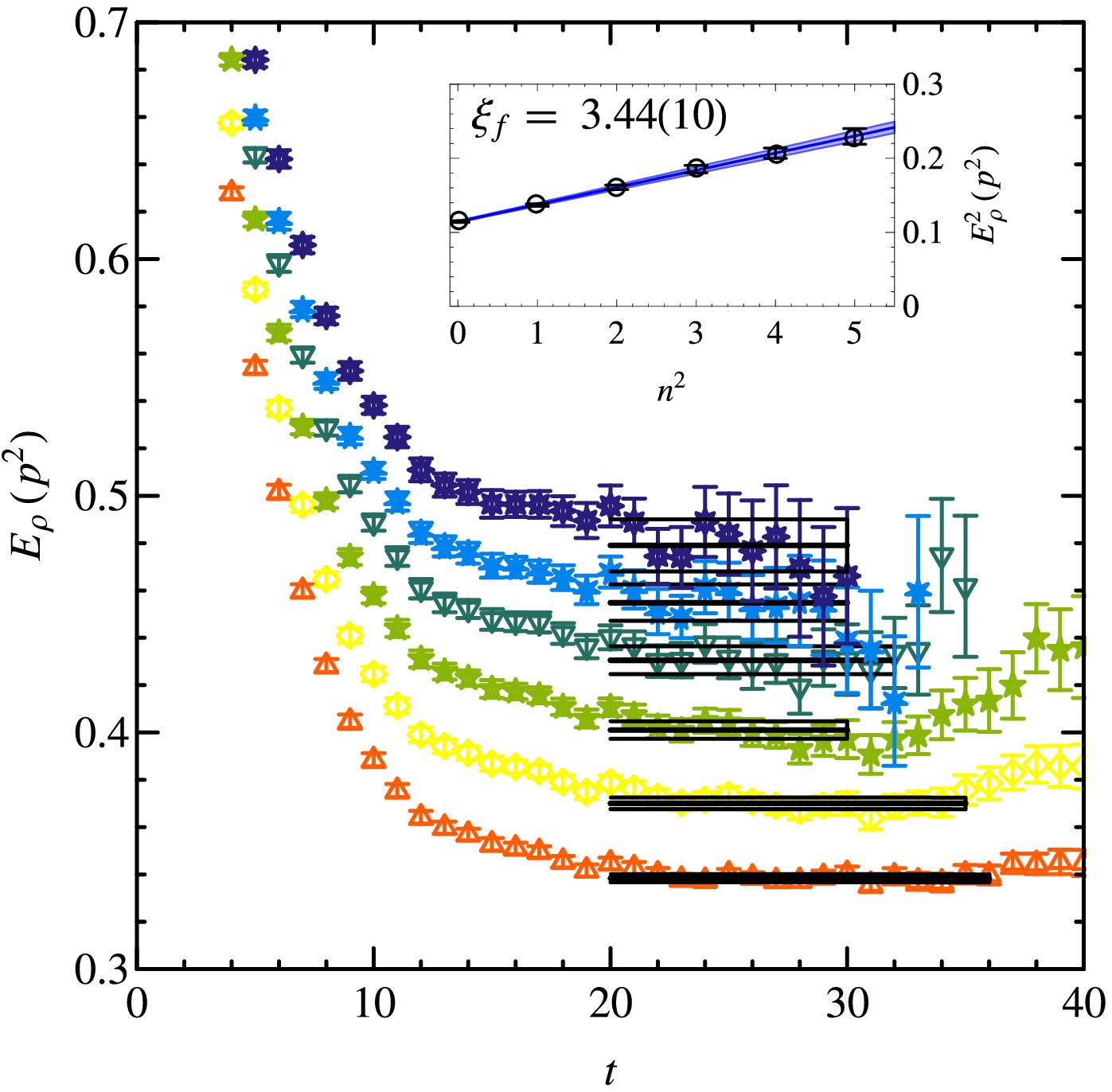}
\caption{Effective mass plots for various momentum projections of the pseudoscalar (left) and vector (right) mesons and corresponding fits. The fermion action parameters for this run are $m_0=-0.0540$, $L_t=96$, $\gamma_g=4.4$ and $\gamma_f=3.3$ with (anti)periodic boundary conditions.
}\label{fig:xif_m054}
\end{figure}

We calculate the energy $E(\vec{p})$ at the spatial momenta
$\vec{p}=2\pi\vec{n}/L_s$ for $n=(0,0,0)$, $(1,0,0)$, $(1,1,0)$, and
$(2,0,0)$. The resulting jackknife energies are used in a linear fit
(in $E(\vec{p})^2$) to Eq.~\ref{eq:disp} to extract $\xi_f$. We find no
significant deviation from linearity. An example is in
Figure~\ref{fig:xif_m054} which shows the pseudoscalar and vector meson
energies with action parameters $\beta=1.5$, $\gamma_g=4.4$,
$m_0=-0.0540$ and $\gamma_f=3.3$ at multiple momentum
projections. (The effective energy is shown as $\ln (C(t+1)/C(t))$.)
We measure $\xi_f=3.44(8)$ from the pseudoscalar meson and
$\xi_f=3.44(10)$ from the vector meson, which are consistent.

In our earlier work, we determined the fermion anisotropy through the
PCAC mass ratio from Schr\"odinger-functional boundary conditions in
the $z$ and $t$ directions. However, unlike the case of the Wilson loop, we found it difficult to exclude excited-state contamination with the limited size of the spacial direction; thus, such a method suffers from a larger systematic error due to the extraction of the PCAC mass. We also
investigated using the Schr\"odinger-functional scheme with Dirichlet
(zero) gauge boundary conditions and a point-like (but smeared)
fermion boundary condition~\cite{Guagnelli:1999zf} in time allowing us
to project our correlators onto non-zero momenta, and thus allowing
an extraction of the fermion anisotropy from the meson dispersion
relation.  This method, however, did not give as reliable results as
in the antiperiodic case.  Therefore, we choose to use the meson
dispersion to determine $\xi_f$ on (anti)periodic boundary gauge
ensembles.

\subsection{Tuning Result}\label{subsec:tuning}

With $\beta$ fixed at $1.5$, we only need to tune the parameters
$\gamma_g$, $\gamma_f$ and $m_0$. Our strategy is to choose some
suitable estimates of the anisotropies, and make a coarse scan in
$m_0$ to find the region of the critical point as determined from the
PCAC mass $M_t$, where, as described in Sec.~\ref{subsec:SF} a
background field in the $t$ direction was used. Due to the nice
properties of the algorithm using the background field, fairly short runs
over roughly 1000 trajectories with measurements every fifth
trajectory are sufficient for the purpose of determining $M_t$. Given
a mass regime of interest, the gauge anisotropy $\xi_g$ is determined
using a background in the $z$ direction as described in
Sec.~\ref{subsec:gauge_aniso}. Again, roughly 1000 trajectories are
sufficient. The measurements of $M_t$ and $\xi_g$ used $12^3\times 32$
volumes, and provide a reasonable range of the bare parameters $m_0$
and $\gamma_g$ for further determinations of the fermion anisotropy
$\xi_f$. These latter measurements used the method described
in Sec.~\ref{subsec:ferm_aniso} on $12^3\times 96$ volumes with antiperiodic boundary conditions in time. Longer runs of roughly 2000 trajectories are used with measurements every fifth trajectory and binned twice, so they are effectively measured every tenth trajectory.

Of course, each of these simulations is independent of the others.
The ensemble parameters, and the results of the determinations of
$M_t$, $\xi_g$, and $\xi_f$ are summarized in Table~\ref{tab:3FsfT},
Table~\ref{tab:3FsfZ}, and Table~\ref{tab:3Fpbc}, respectively.

\begin{figure}
\includegraphics[angle=0,width=0.7\textwidth]{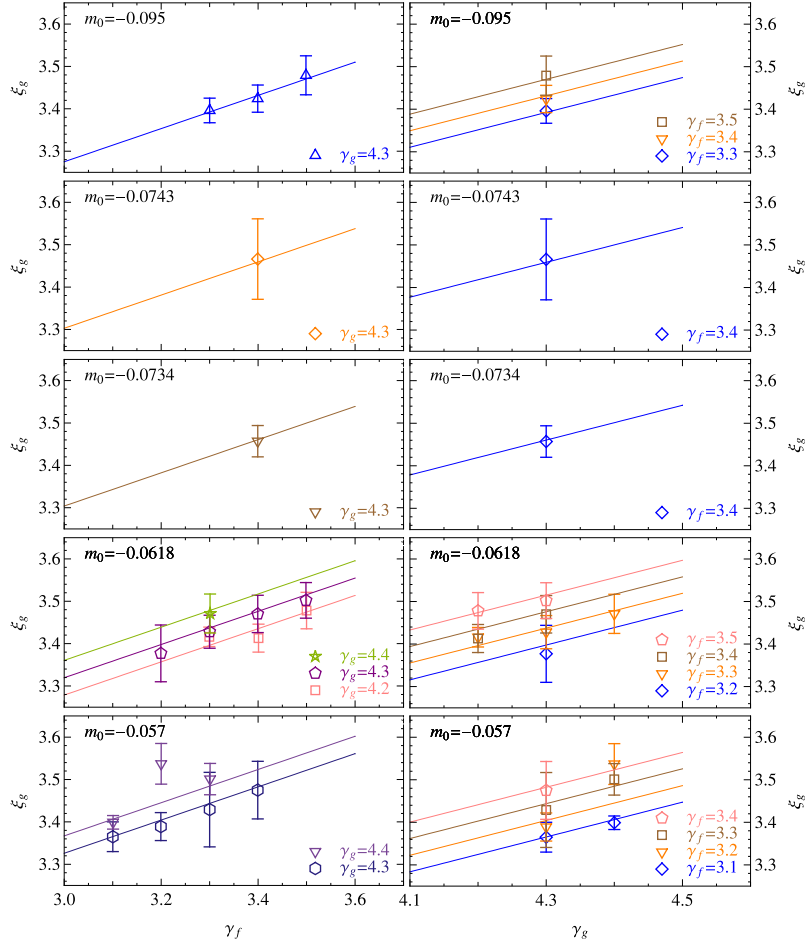}
\caption{Gauge anisotropy as a function of $\gamma_f$, $\gamma_g$ at
various fixed $m_0$. The straight lines are the fit functions in
Eq.~\ref{eq:3conditions} keeping the remaining two parameters
fixed. Table~\ref{tab:3FsfZ} lists the details of the ensemble
parameters.
}
\label{fig:xig-cubefit}
\end{figure}

\begin{figure}
\includegraphics[angle=0,width=0.7\textwidth]{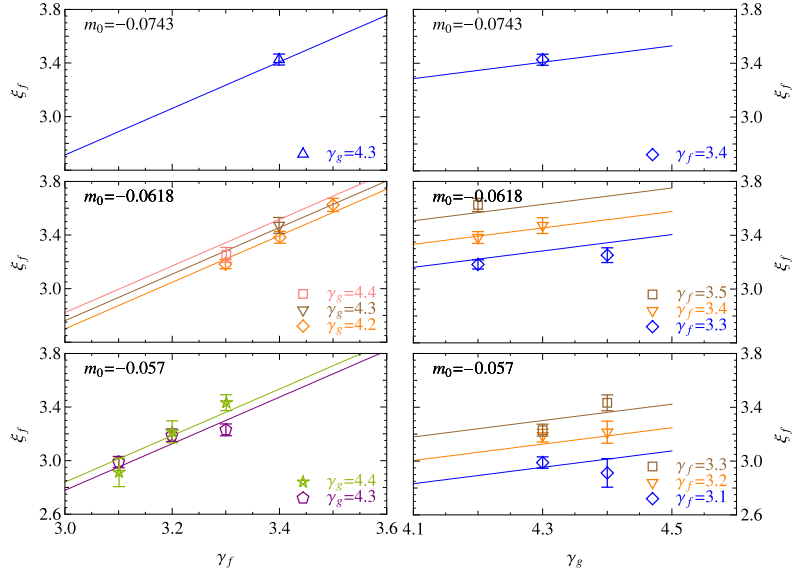}
\caption{Fermion anisotropy as a function of $\gamma_f$, $\gamma_g$ at
various fixed $m_0$. Table~\ref{tab:3Fpbc} lists the details of the ensemble
parameters.
}\label{fig:xif-cubefit}
\end{figure}

\begin{figure}
\includegraphics[angle=0,width=0.7\textwidth]{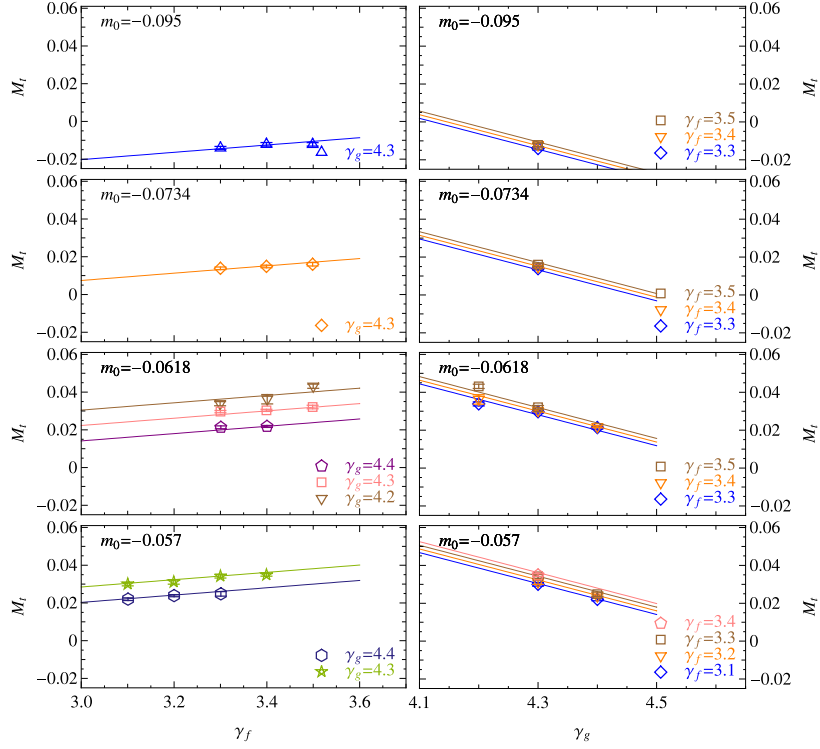}
\caption{PCAC mass (in units of the temporal lattice spacing) as a
function of $\gamma_f$, $\gamma_g$ at various fixed $m_0$. Table~\ref{tab:3FsfT} lists the details of the ensemble
parameters.
}\label{fig:pcac-cubefit}
\end{figure}

\begin{figure}
\includegraphics[angle=0,width=0.7\textwidth]{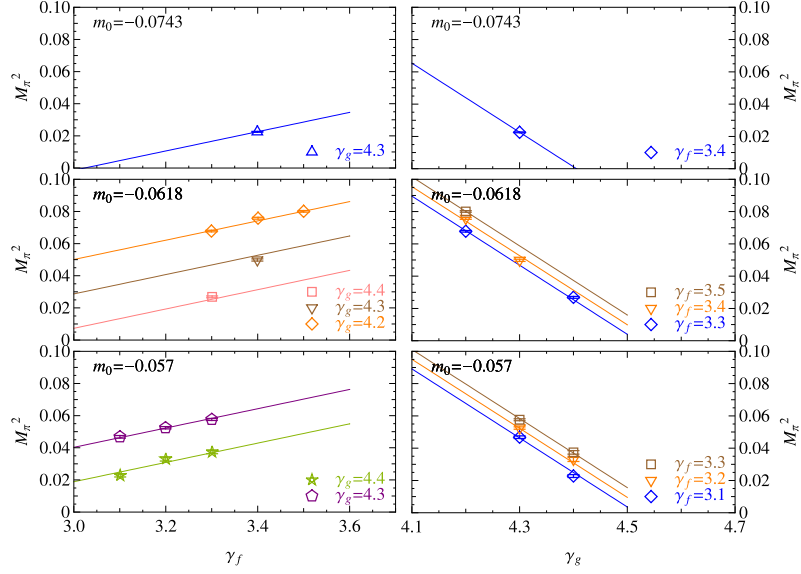}
\caption{Pion mass squared (in units of $a_t^{-2}$) as a function of $\gamma_f$, $\gamma_g$ at various fixed $m_0$. Table~\ref{tab:3Fpbc} lists the details of the ensemble parameters.
}\label{fig:mpisq-cubefit}
\end{figure}

\begin{figure}
\includegraphics[angle=0,width=0.7\textwidth]{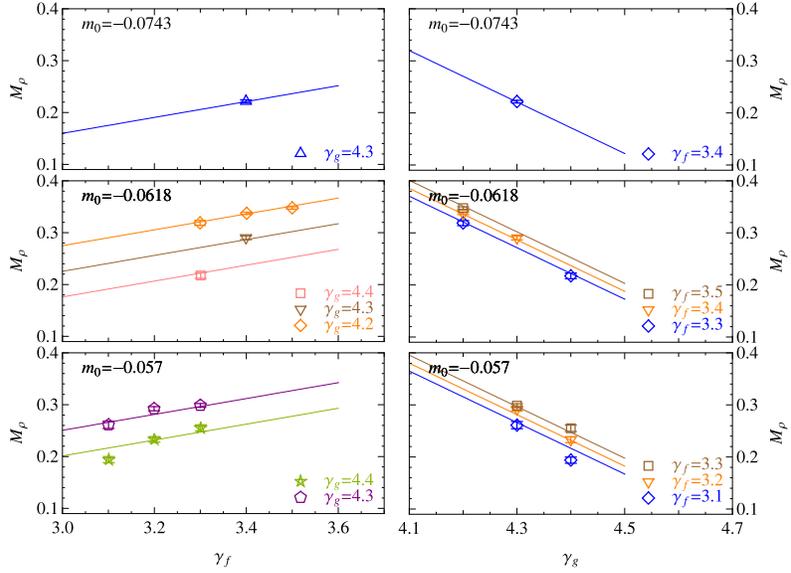}
\caption{Rho-meson mass (in units of $a_t^{-1}$) as a function of $\gamma_f$, $\gamma_g$ at various fixed $m_0$. Table~\ref{tab:3Fpbc} lists the details of the ensemble parameters.
}\label{fig:mrho-cubefit}
\end{figure}

For each measurement, we fit the data according to the linear ansatz
in Eq.~\ref{eq:3conditions}. The fit parameters are
\begin{alignat}{4}
\label{eq:cube-fit_param}
a_0 &= 0.5(8), &\quad a_1 &= 0.41(14), &\quad a_2 &= 0.39(8), &\quad a_3 &= 1.3(7) \nonumber \\
b_0 &= -4.8(12), &\quad b_1 &= 0.6(2), &\quad b_2 &= 1.74(17), &\quad b_3 &= 4(3) \nonumber \\
c_0 &= 0.395(18), &\quad c_1 &= -0.082(3), &\quad c_2 &= 0.0194(17), &\quad c_3 &= 1.282(16).
\end{alignat}
The values of $\chi^2$ per degree of freedom are $6.0/16$, $13.2/8$,
$34.7/17$ respectively. Figures~\ref{fig:xig-cubefit},
\ref{fig:xif-cubefit} and \ref{fig:pcac-cubefit} demonstrate how the
fits work using two-dimensional slices of the parameter and
measurement spaces. Both $\xi_g$ and $\xi_f$ show little sensitivity to
the parameter $m_0$, and thus the slope is poorly determined in our
fits. However, we do not expect this would affect our final result
much since the changes in $m_0$ are  $O(10^{-2})$ and the $\xi_g$ and $\xi_f$ are  $O(1)$. The interactions of $\xi_g$ and $\xi_f$ also have large uncertainties in the fits; this is also expected since the running coefficients, $\gamma_f$ and $\gamma_g$, are around 3--4 and there is a long extrapolation to 0. In terms of fits to $M_t$, $m_0$ has the
dominant effect, a few times larger than $\gamma_g$ while
$\gamma_f$ has one magnitude smaller contribution. We found that $\xi_g$ and $\xi_f$  have positive linear dependence on the bare parameters, while for the case of $M_t$ versus $\gamma_g$, we found that $M_t$ increases while $\gamma_g$ decreases. The $\gamma_g$ parameter is the dominant factor in $\xi=a_s/a_t$. Increasing it, $a_t$ will increase as well; this leads to the only non-dimensionless measurement $M_t$ becoming smaller in units of $a_t$. Similar effects are observed when we parameterize the pion mass squared or rho-meson mass (which are measured on the antiperiodic
boundary condition ensemble); see Figure~\ref{fig:mpisq-cubefit} and
\ref{fig:mrho-cubefit}.

We can gain a qualitative understanding of the origin of the opposite sign in the slopes of $\gamma_g$ versus $\gamma_f$ for the mass measurements
by considering the classical dispersion relation for the action in
Eq.~\ref{eq:fermion-action}.  For a small background chromo-electric
field $F_{0i}$, we find that the lattice dispersion relation energy at zero spatial momentum $\hat{E}=2\sinh(a_t m_0)/a_t$ satisfies for our
choice of clover coefficient $c_t$ from Eq.~\ref{eq:clov_coeffs}:
\begin{eqnarray}\label{eq:disp_rel}
\hat{E}^2 (1 + a_t m_0) = m_0^2 -
\left(\frac{\gamma_g}{\gamma_f} +
\frac{1}{2}\left(\xi \frac{\gamma_g}{\gamma_f} + 1\right)\sinh(a_t m_0)\right)\sum_i \sigma_{0i} F_{0i}.
\end{eqnarray}
Thus, the derivatives $\partial \hat{E}^2/{\partial\gamma_g} \propto -F_{0i}$ and
$\partial \hat{E}^2/{\partial\gamma_f} \propto +F_{0i}$ have opposite sign.

\begin{figure}
\includegraphics[angle=0,width=0.5\textwidth]{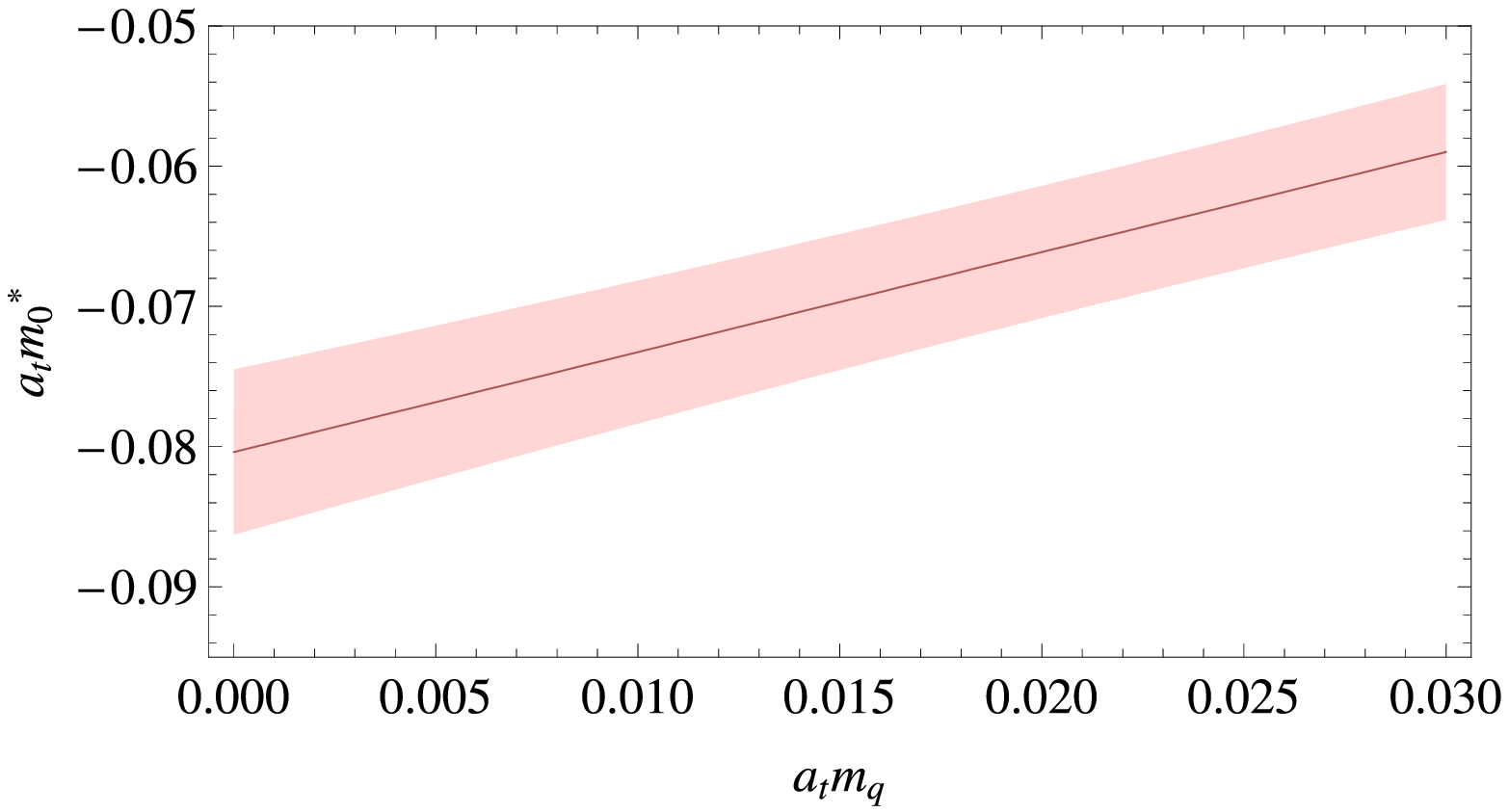}
\includegraphics[angle=0,width=0.5\textwidth]{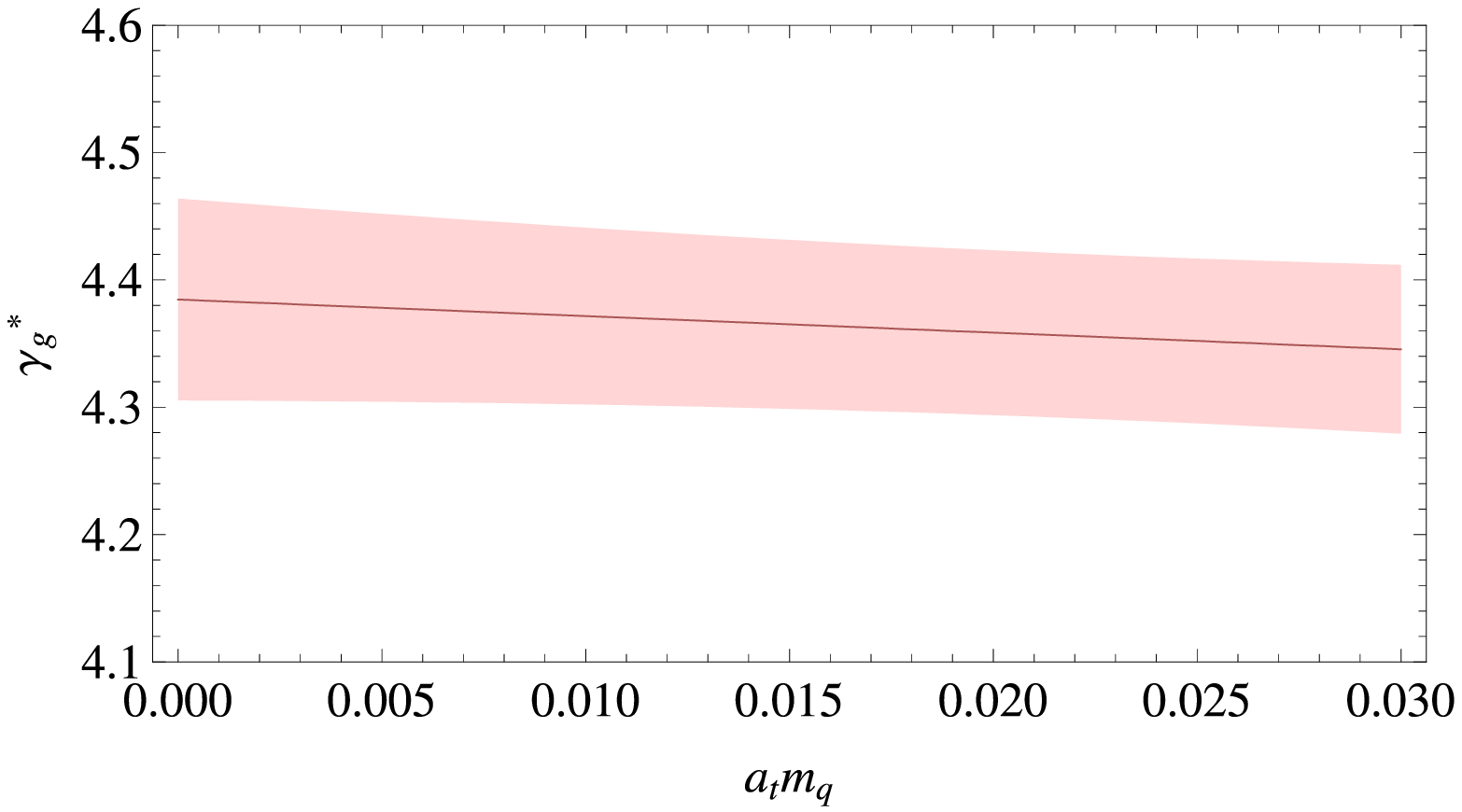}
\includegraphics[angle=0,width=0.5\textwidth]{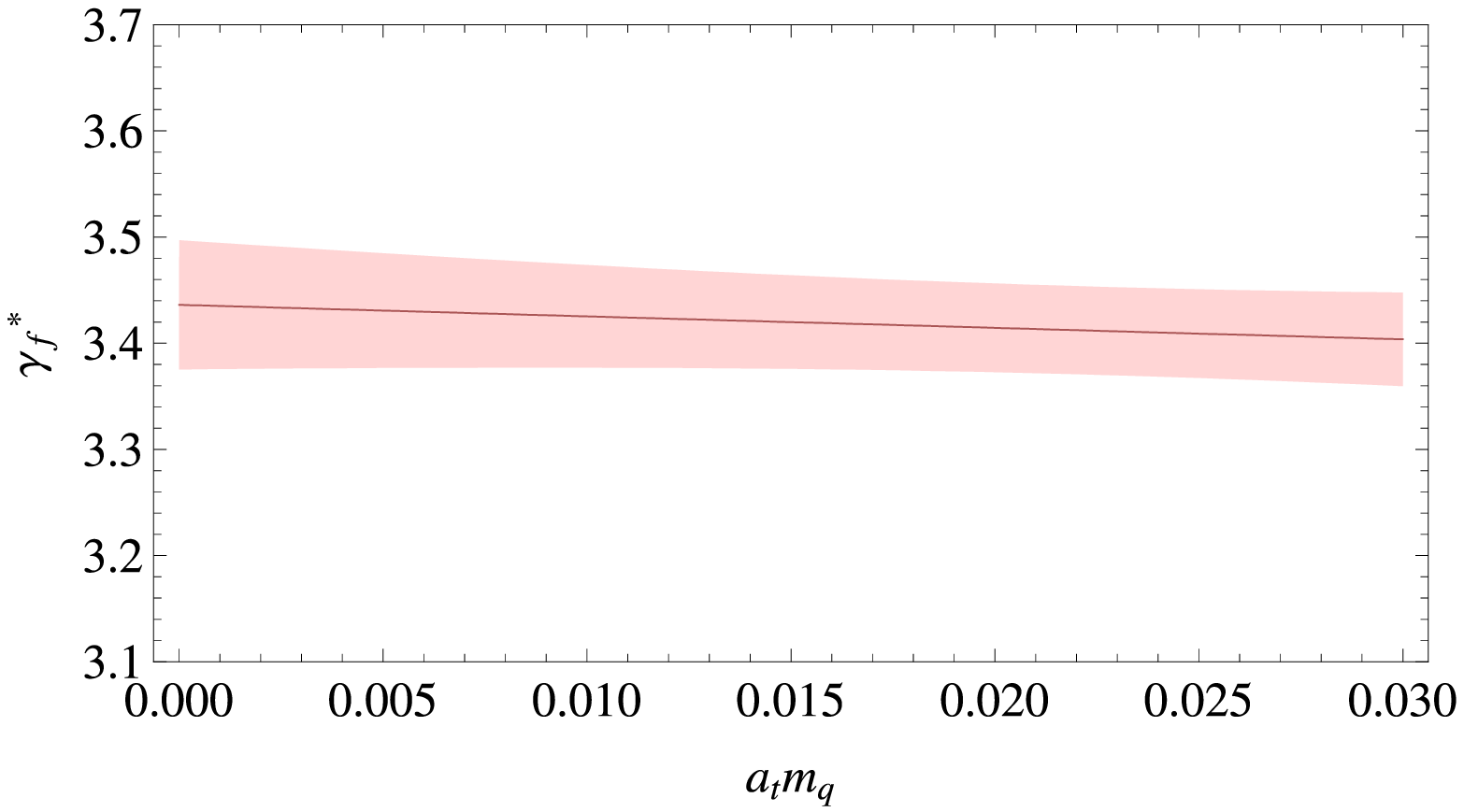}
\caption{Tuned gauge and fermion action parameters, $m_0^*$ (top),
$\gamma_g^*$ (middle) and $\gamma_f^*$ (bottom), as functions of the
bare quark mass, $a_t m_q$, corresponding to the solution
of Eq.~\ref{eq:solver}. There is negligible mass dependence in the
tuned anisotropy parameters. With a spatial lattice spacing
$a_s=0.12$~fm, the maximum extent of the horizontal axis is about
$175$~MeV.}
\label{fig:FermAct-mq}
\end{figure}

We impose the renormalization condition that in the chiral limit
$\{\xi_g,\xi_f,M_t\}=\{3.5,3.5,0\}$. By solving Eq.~\ref{eq:solver}
with the parameters in Eq.~\ref{eq:cube-fit_param}, we found that
\begin{eqnarray}\label{eq:predicted-par}
\{m_{cr},\gamma_g^*,\gamma_f^*\}=\{-0.080(6), 4.38(8), 3.44(7)\}.
\end{eqnarray}
By fixing the renormalization condition in the general case,
$\{\xi_g,\xi_f,M_t\}=\{3.5,3.5,m_q\}$, we obtain the desired action
parameters as a functions of the bare quark mass, as shown in
Fig.~\ref{fig:FermAct-mq}. As emphasized in the discussions above,
$\gamma_g^*$ and $\gamma_f^*$ have very small quark mass dependence
from the chiral limit up to the heaviest $m_0$ used in this work,
$-0.057$ (which corresponds to $m_q\approx 0.03$). Of course, $m_0$ is
linearly proportional to the bare quark mass.

\begin{figure}
\includegraphics[angle=0,width=0.45\textwidth]{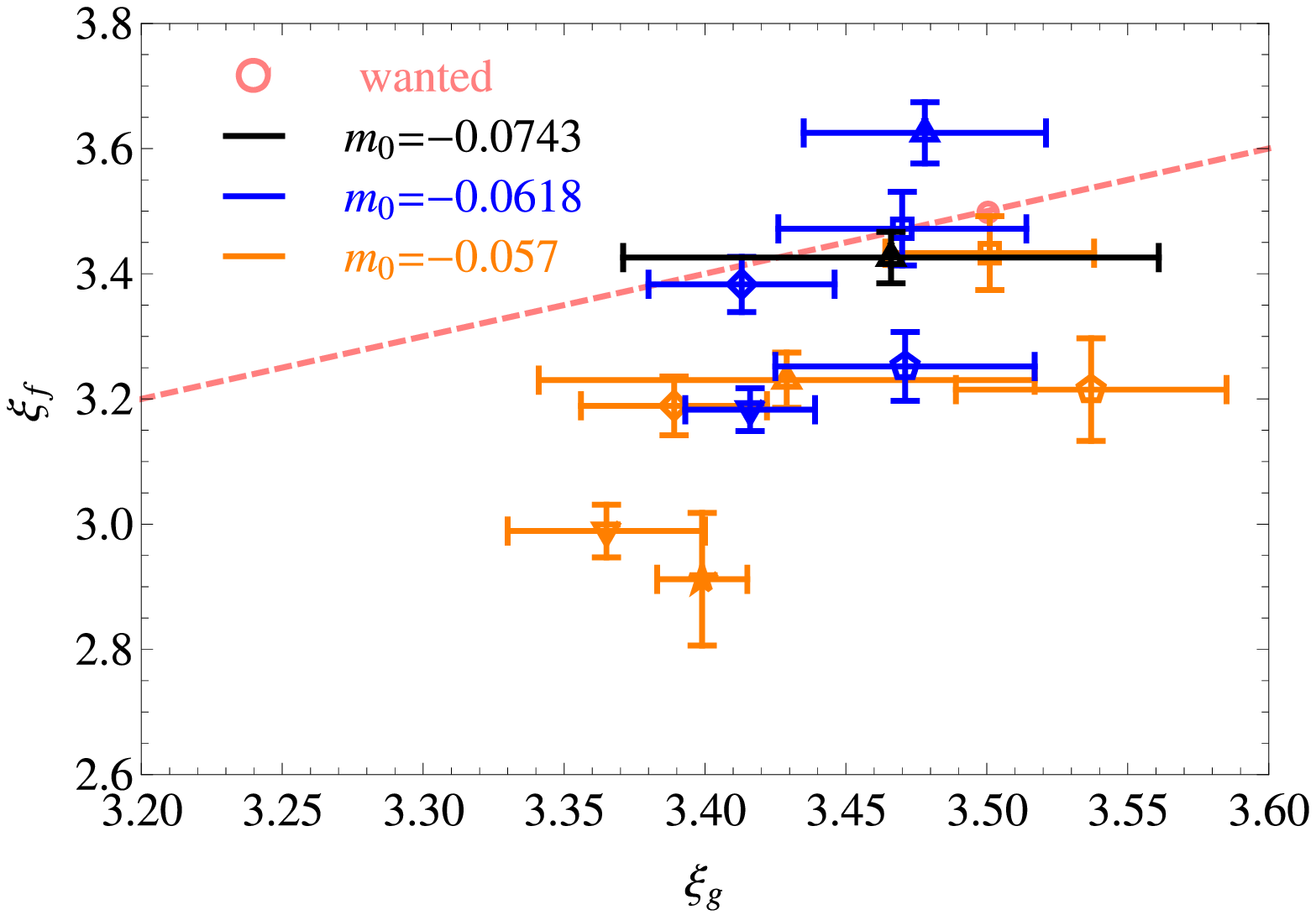}
\includegraphics[angle=0,width=0.45\textwidth]{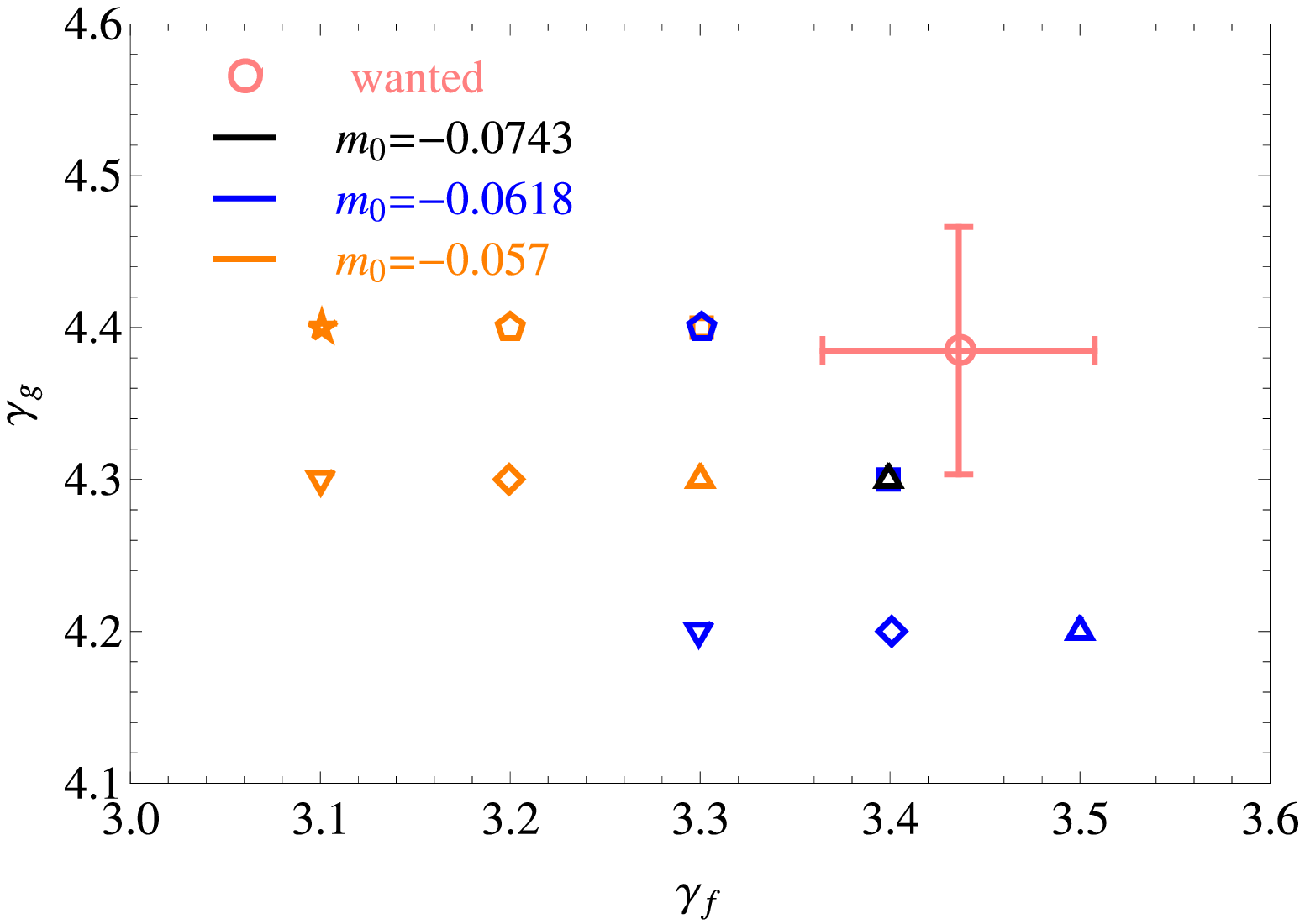}
\caption{Renormalized gauge and fermion anisotropies (left) and the
corresponding bare parameter (left). The detailed parameters can be
found in Table~\ref{tab:3Fpbc} and Table~\ref{tab:3FsfZ}. Note that
the predicted coefficients from fitting Eq.~\ref{eq:solver} in the
chiral limit are $\{\gamma_g^*,\gamma_f^*\}=\{4.38(8), 3.44(7)\}$.}
\label{fig:joint_renom}
\end{figure}

Figure~\ref{fig:joint_renom} shows a subset of the $m_0$ values with
their corresponding gauge and fermion anisotropies. The two simulation
points that are consistent with the desired points of
$\{\xi_g,\xi_f\}=\{3.5,3.5\}$ are consistent with
Eq.~\ref{eq:predicted-par}.

\begin{figure}
\includegraphics[angle=0,width=0.5\textwidth]{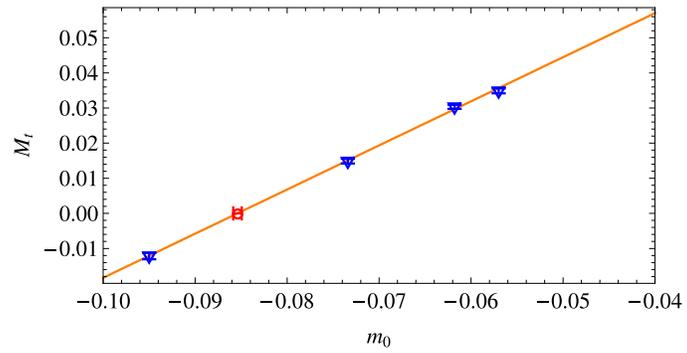}
\caption{PCAC mass at $\beta=1.5$, $\gamma_g=4.3$ and $\gamma_f=3.4$
as a function of $m_0$. A linear interpolation gives
$m_{\rm cr}=-0.0854(5)$.}
\label{fig:pcac-m0}
\end{figure}

Based upon the negligible mass dependence that is observed, we henceforth
fix the anisotropy bare parameters to
\begin{equation}
\gamma_g^* = 4.3, \quad \gamma_f^* = 3.4
\end{equation}
which corresponds to a parameter set used in Figure~\ref{fig:joint_renom}.
The corresponding clover coefficients are
\begin{eqnarray*}
c_s = 1.589, \quad
c_t = 0.903.
\end{eqnarray*}
Figure~\ref{fig:pcac-m0} shows a PCAC measurement as a function of
$m_0$ with $\beta=1.5$, $\gamma_g=4.4$, and $\gamma_f=3.3$; and linear
interpolation to $M_t=0$ gives us
\begin{equation}
a_t m_{cr} = -0.0854(5)\quad .
\end{equation}

\begin{figure}
\includegraphics[width=0.5\textwidth]{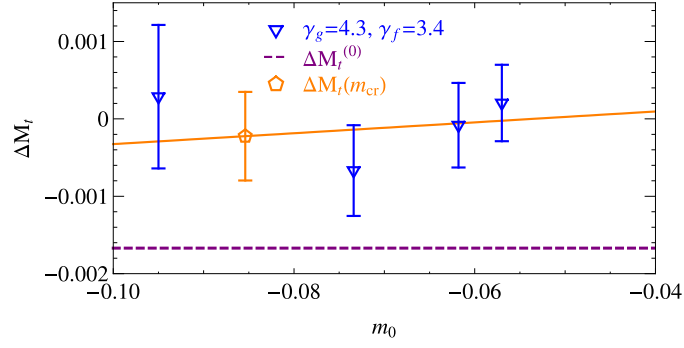}
\caption{The measured $\Delta M_{t}$ (down triangles) in units of
$a_t$ as a function of $m_0$ at fixed $\gamma_g=4.3$ and
$\gamma_f=3.4$, along with $\Delta M_{t}^{(0)}$ (dashed line)
}\label{fig:dmt_vs_tree}
\end{figure}

Finally, we will discuss how our initial tadpole-improved clover
coefficients $c_{\rm s,t}$ differ from the nonperturbative
coefficients in the Schr\"odinger-functional scheme (as described in
Sec.~\ref{subsec:SF}). Table~\ref{tab:3FsfT} lists the PCAC mass,
$\Delta M_t$ and its tree-level value (in units of $a_t^{-1}$)
measured with the $t$-direction Schr\"odinger-functional boundary
condition. If we had nonperturbatively tuned $c_{s,t}$ and imposed the
renormalization conditions from Eq.~\ref{eq:5conditions}, we should
expect that $\Delta M_t=\Delta M_t^{(0)}$. There are four sets of
ensemble parameters that satisfy the conditions on $\gamma_g^*$ and
$\gamma_f^*$. If we extrapolate $\Delta M_t$ to $m_{cr}$, we find
$\Delta M_t = -0.00022(57)$, which is about 1.5 standard deviations
away from the tree-level value: $-0.00167$. (See
Figure~\ref{fig:dmt_vs_tree}.) We conclude that the tadpole-corrected
tree-level coefficients with stout-link smearing are close enough to
the nonperturbative $O(a)$-improved coefficients in the three-flavor
dynamical simulation.

\begin{table}
\caption{\label{tab:3FsfT} PCAC mass (in units of the inverse temporal lattice spacing)
measured from the $t$-direction Schr\"odinger functional boundary
condition ensemble with volume $12^3\times 32$.}
\begin{center}
\begin{tabular}{ccc|ccc}
\hline
\hline
$m_0$ & $\gamma_g$ & $\gamma_f$ & $M_t$ & $\Delta M_t$ & $\Delta M_t^{(0)}$ \\
\hline
 $-$0.0950 & 4.3 & 3.5 & $-$0.0122(9) &  0.0003(10) & $-$0.001547 \\
 $-$0.0950 & 4.3 & 3.4 & $-$0.0121(9) &  0.0003(9) & $-$0.00167 \\
 $-$0.0950 & 4.3 & 3.3 & $-$0.0141(8) &  0.0002(9) & $-$0.001798 \\
 $-$0.0734 & 4.3 & 3.5 &  0.0160(9) & $-$0.0007(9) & $-$0.001547 \\
 $-$0.0734 & 4.3 & 3.4 &  0.0149(7) & $-$0.0007(6) & $-$0.00167 \\
 $-$0.0734 & 4.3 & 3.3 &  0.0139(7) &  0.0006(11) & $-$0.001798 \\
 $-$0.0618 & 4.2 & 3.5 &  0.0431(11) &  0.0002(4) & $-$0.001427 \\
 $-$0.0618 & 4.2 & 3.4 &  0.036(2) & $-$0.0004(8) & $-$0.001545 \\
 $-$0.0618 & 4.2 & 3.3 &  0.0339(11) &  0.0004(8) & $-$0.001672 \\
 $-$0.0618 & 4.3 & 3.5 &  0.0321(9) &  0.0003(5) & $-$0.001547 \\
 $-$0.0618 & 4.3 & 3.4 &  0.0303(6) & $-$0.0001(5) & $-$0.00167 \\
 $-$0.0618 & 4.3 & 3.3 &  0.0297(6) & $-$0.0003(4) & $-$0.001798 \\
 $-$0.0618 & 4.4 & 3.4 &  0.0218(5) & $-$0.0004(5) & $-$0.001791 \\
 $-$0.0618 & 4.4 & 3.3 &  0.0213(7) &  0.0001(6) & $-$0.001923 \\
 $-$0.0570 & 4.3 & 3.4 &  0.0349(7) &  0.0002(5) & $-$0.00167 \\
 $-$0.0570 & 4.3 & 3.3 &  0.0342(10) &  0.0004(7) & $-$0.001798 \\
 $-$0.0570 & 4.3 & 3.2 &  0.0311(12) &  0.0005(8) & $-$0.001935 \\
 $-$0.0570 & 4.3 & 3.1 &  0.0300(9) &  0.0025(8) & $-$0.002078 \\
 $-$0.0570 & 4.4 & 3.3 &  0.0248(12) &  0.0002(8) & $-$0.001923 \\
 $-$0.0570 & 4.4 & 3.2 &  0.0239(6) &  0.0005(7) & $-$0.002065 \\
 $-$0.0570 & 4.4 & 3.1 &  0.0220(7) &  0.0012(11) & $-$0.002212 \\
\hline
\hline
\end{tabular}
\end{center}
\end{table}

\begin{table}
\caption{\label{tab:3FsfZ} Renormalized gauge anisotropy measured from the $z$-direction
Schr\"odinger functional boundary condition ensemble with volume
$12^3\times 32$.}
\begin{center}
\begin{tabular}{ccc|c}
\hline\hline
$m_0$ & $\gamma_g$ & $\gamma_f$ & $\xi_g$ \\
\hline
 $-$0.0950 & 4.3 & 3.5 &  3.48(5) \\
 $-$0.0950 & 4.3 & 3.4 &  3.42(3) \\
 $-$0.0950 & 4.3 & 3.3 &  3.40(3) \\
 $-$0.0743 & 4.3 & 3.4 &  3.47(10) \\
 $-$0.0734 & 4.3 & 3.4 &  3.46(4) \\
 $-$0.0618 & 4.2 & 3.5 &  3.48(4) \\
 $-$0.0618 & 4.2 & 3.4 &  3.41(3) \\
 $-$0.0618 & 4.2 & 3.3 &  3.42(2) \\
 $-$0.0618 & 4.3 & 3.5 &  3.50(4) \\
 $-$0.0618 & 4.3 & 3.4 &  3.47(4) \\
 $-$0.0618 & 4.3 & 3.3 &  3.43(4) \\
 $-$0.0618 & 4.3 & 3.2 &  3.38(7) \\
 $-$0.0618 & 4.4 & 3.3 &  3.47(5) \\
 $-$0.0570 & 4.3 & 3.4 &  3.48(7) \\
 $-$0.0570 & 4.3 & 3.3 &  3.43(9) \\
 $-$0.0570 & 4.3 & 3.2 &  3.39(3) \\
 $-$0.0570 & 4.3 & 3.1 &  3.36(4) \\
 $-$0.0570 & 4.4 & 3.3 &  3.50(4) \\
 $-$0.0570 & 4.4 & 3.2 &  3.54(5) \\
 $-$0.0570 & 4.4 & 3.1 &  3.399(16) \\
\hline\hline
\end{tabular}
\end{center}
\end{table}

\begin{table}
\caption{\label{tab:3Fpbc} Fermion anisotropy, vector-meson mass and pseudoscalar mass
squared (in units of the inverse temporal lattice spacing) from the
periodic boundary condition ensemble with volume $12^3\times 96$.}
\begin{center}
\begin{tabular}{ccc|cccc}
\hline
\hline
$m_0$ & $\gamma_g$ & $\gamma_f$ & $\xi_f$ & $m_\pi$ & $m_\rho$ & $m_\pi/m_\rho$ \\
\hline
 $-$0.0743 & 4.3 & 3.4 &  3.43(4) &  0.1501(9) &  0.222(3) &  0.677(9) \\
 $-$0.0618 & 4.2 & 3.5 &  3.62(5) &  0.2830(9) &  0.348(2) &  0.814(5) \\
 $-$0.0618 & 4.2 & 3.4 &  3.38(4) &  0.2753(10) &  0.337(2) &  0.816(5) \\
 $-$0.0618 & 4.2 & 3.3 &  3.18(3) &  0.2604(10) &  0.319(5) &  0.817(11) \\
 $-$0.0618 & 4.3 & 3.4 &  3.47(6) &  0.2232(15) &  0.290(4) &  0.769(9) \\
 $-$0.0618 & 4.4 & 3.3 &  3.25(6) &  0.1639(17) &  0.217(5) &  0.754(16) \\
 $-$0.0570 & 4.3 & 3.3 &  3.23(4) &  0.2401(13) &  0.299(4) &  0.804(8) \\
 $-$0.0570 & 4.3 & 3.2 &  3.19(5) &  0.2290(16) &  0.292(4) &  0.784(10) \\
 $-$0.0570 & 4.3 & 3.1 &  2.99(4) &  0.2164(18) &  0.261(7) &  0.828(20) \\
 $-$0.0570 & 4.4 & 3.3 &  3.43(6) &  0.193(3) &  0.255(6) &  0.758(17) \\
 $-$0.0570 & 4.4 & 3.2 &  3.22(8) &  0.182(3) &  0.233(5) &  0.780(17) \\
 $-$0.0570 & 4.4 & 3.1 &  2.91(11) &  0.151(3) &  0.194(6) &  0.78(2) \\
\hline
\hline
\end{tabular}
\end{center}
\end{table}

\section{Conclusion and Outlook}\label{Sec:Conclusion}

This is the first calculation that combines Schr\"odinger functional
with stout-link smearing on three-flavor anisotropic clover
action. The Schr\"odinger-functional boundary conditions allow us to work on small volumes and small quark masses with improved signal, which
improves our ability to probe the chiral limit. Stout-link smearing
improves the chiral and scaling properties of our chirally broken
fermion action. We take advantage of the much smaller cost by
implementing both in our calculation.

We studied a range of stout-link parameters that may be safely applied
to multiple choices of $\beta$ and different coefficients in the
fermion sector in three-flavor simulation. Although our numerical
simulations suggested that higher values of $\rho$ could be applied,
in the end, the stout-link parameters we chose conservatively to
be $\rho=0.14$ and $n_\rho=2$, as suggested by a one-loop perturbative
calculation.

In a preliminary three-flavor study of various $\beta$ values, we
found through the static quark potential that $\beta=1.5$ gave us the
desired spatial lattice spacing around $0.12$~fm in the physical limit.
The remaining coefficients in the gauge action and clover coefficients
$c_{s,t}$ in the fermion action were set to their tree-level tadpole improved
values, which were numerically determined and interpolated using a
Pad\'e approximation.

We determined the gauge anisotropy using Wilson-loop ratios and
Schr\"odinger-functional boundary conditions in the $z$ direction; we
determined the fermion anisotropy using antiperiodic
boundary conditions in time. The quark mass was estimated from PCAC
measurements using Schr\"odinger-functional boundary conditions in the $t$ direction from which we found the critical mass.

We then tuned the remaining parameters $\xi_{g,f}$ to achieve the desired
renormalized anisotropy $\xi=3.5$ in the chiral limit. We found
$\{m_{cr},\gamma_g^*,\gamma_f^*\}=\{-0.080(6), 4.38(8), 3.44(7)\}$
from a linear parameterization. The mass dependence of the gauge and
fermion anisotropies is found to be negligible. Hence, we have for the
final parameters
\begin{eqnarray*}
\gamma_g^* = 4.3, \quad \gamma_f^* = 3.4, \quad
c_s = 1.589, \quad
c_t = 0.903,
\end{eqnarray*}
with $a_tm_{cr} = -0.0854(5)$. Further, we showed in the Schr\"odinger-functional scheme that when using
stout-link smearing and numerically determined tadpole factors, our
fermion action automatically fulfills an (on-shell) $O(a)$-improved
renormalization condition. In particular, the clover coefficients $c_{s,t}$ are consistent with being nonperturbatively tuned.
We will apply the same approach for future tuning as
we move to finer lattices.

With the determined coefficients from this work, we are currently
generating $N_f=2+1$-flavor ensembles with multiple masses. Further
measurements are being made during the gauge generation process to
precisely determine the lattice spacing in the chiral limit and some
hadronic properties~\cite{Edwards:aniso_tune}.

\section*{Acknowledgements}
We thank Michael Peardon for several valuable discussions.
This work was done using the Chroma software
suite~\cite{Edwards:2004sx} on clusters at Jefferson Laboratory using
time awarded under the USQCD Initiative. Authored by Jefferson
Science Associates, LLC under U.S. DOE Contract
No. DE-AC05-06OR23177. The U.S. Government retains a non-exclusive,
paid-up, irrevocable, world-wide license to publish or reproduce this
manuscript for U.S. Government purposes.



\begin{thebibliography}{49}
\expandafter\ifx\csname natexlab\endcsname\relax\def\natexlab#1{#1}\fi
\expandafter\ifx\csname bibnamefont\endcsname\relax
  \def\bibnamefont#1{#1}\fi
\expandafter\ifx\csname bibfnamefont\endcsname\relax
  \def\bibfnamefont#1{#1}\fi
\expandafter\ifx\csname citenamefont\endcsname\relax
  \def\citenamefont#1{#1}\fi
\expandafter\ifx\csname url\endcsname\relax
  \def\url#1{\texttt{#1}}\fi
\expandafter\ifx\csname urlprefix\endcsname\relax\def\urlprefix{URL }\fi
\providecommand{\bibinfo}[2]{#2}
\providecommand{\eprint}[2][]{\url{#2}}

\bibitem[{\citenamefont{Sasaki et~al.}(2002)\citenamefont{Sasaki, Blum, and
  Ohta}}]{Sasaki:2001nf}
\bibinfo{author}{\bibfnamefont{S.}~\bibnamefont{Sasaki}},
  \bibinfo{author}{\bibfnamefont{T.}~\bibnamefont{Blum}}, \bibnamefont{and}
  \bibinfo{author}{\bibfnamefont{S.}~\bibnamefont{Ohta}},
  \bibinfo{journal}{Phys. Rev.} \textbf{\bibinfo{volume}{D65}},
  \bibinfo{pages}{074503} (\bibinfo{year}{2002}), \eprint{hep-lat/0102010}.

\bibitem[{\citenamefont{Guadagnoli et~al.}(2004)\citenamefont{Guadagnoli,
  Papinutto, and Simula}}]{Guadagnoli:2004wm}
\bibinfo{author}{\bibfnamefont{D.}~\bibnamefont{Guadagnoli}},
  \bibinfo{author}{\bibfnamefont{M.}~\bibnamefont{Papinutto}},
  \bibnamefont{and} \bibinfo{author}{\bibfnamefont{S.}~\bibnamefont{Simula}},
  \bibinfo{journal}{Phys. Lett.} \textbf{\bibinfo{volume}{B604}},
  \bibinfo{pages}{74} (\bibinfo{year}{2004}), \eprint{hep-lat/0409011}.

\bibitem[{\citenamefont{Leinweber et~al.}(2005)\citenamefont{Leinweber,
  Melnitchouk, Richards, Williams, and Zanotti}}]{Leinweber:2004it}
\bibinfo{author}{\bibfnamefont{D.~B.} \bibnamefont{Leinweber}},
  \bibinfo{author}{\bibfnamefont{W.}~\bibnamefont{Melnitchouk}},
  \bibinfo{author}{\bibfnamefont{D.~G.} \bibnamefont{Richards}},
  \bibinfo{author}{\bibfnamefont{A.~G.} \bibnamefont{Williams}},
  \bibnamefont{and} \bibinfo{author}{\bibfnamefont{J.~M.}
  \bibnamefont{Zanotti}}, \bibinfo{journal}{Lect. Notes Phys.}
  \textbf{\bibinfo{volume}{663}}, \bibinfo{pages}{71} (\bibinfo{year}{2005}),
  \eprint{nucl-th/0406032}.

\bibitem[{\citenamefont{Sasaki et~al.}(2005)\citenamefont{Sasaki, Sasaki, and
  Hatsuda}}]{Sasaki:2005ap}
\bibinfo{author}{\bibfnamefont{K.}~\bibnamefont{Sasaki}},
  \bibinfo{author}{\bibfnamefont{S.}~\bibnamefont{Sasaki}}, \bibnamefont{and}
  \bibinfo{author}{\bibfnamefont{T.}~\bibnamefont{Hatsuda}},
  \bibinfo{journal}{Phys. Lett.} \textbf{\bibinfo{volume}{B623}},
  \bibinfo{pages}{208} (\bibinfo{year}{2005}), \eprint{hep-lat/0504020}.

\bibitem[{\citenamefont{Sasaki and Sasaki}(2005)}]{Sasaki:2005ug}
\bibinfo{author}{\bibfnamefont{K.}~\bibnamefont{Sasaki}} \bibnamefont{and}
  \bibinfo{author}{\bibfnamefont{S.}~\bibnamefont{Sasaki}},
  \bibinfo{journal}{Phys. Rev.} \textbf{\bibinfo{volume}{D72}},
  \bibinfo{pages}{034502} (\bibinfo{year}{2005}), \eprint{hep-lat/0503026}.

\bibitem[{\citenamefont{Burch et~al.}(2006)}]{Burch:2006cc}
\bibinfo{author}{\bibfnamefont{T.}~\bibnamefont{Burch}} \bibnamefont{et~al.},
  \bibinfo{journal}{Phys. Rev.} \textbf{\bibinfo{volume}{D74}},
  \bibinfo{pages}{014504} (\bibinfo{year}{2006}), \eprint{hep-lat/0604019}.

\bibitem[{\citenamefont{Mathur et~al.}(2005)}]{Mathur:2003zf}
\bibinfo{author}{\bibfnamefont{N.}~\bibnamefont{Mathur}} \bibnamefont{et~al.},
  \bibinfo{journal}{Phys. Lett.} \textbf{\bibinfo{volume}{B605}},
  \bibinfo{pages}{137} (\bibinfo{year}{2005}), \eprint{hep-ph/0306199}.

\bibitem[{\citenamefont{Chen}(2001)}]{Chen:2000ej}
\bibinfo{author}{\bibfnamefont{P.}~\bibnamefont{Chen}}, \bibinfo{journal}{Phys.
  Rev.} \textbf{\bibinfo{volume}{D64}}, \bibinfo{pages}{034509}
  (\bibinfo{year}{2001}), \eprint{hep-lat/0006019}.

\bibitem[{\citenamefont{Okamoto et~al.}(2002)}]{Okamoto:2001jb}
\bibinfo{author}{\bibfnamefont{M.}~\bibnamefont{Okamoto}} \bibnamefont{et~al.}
  (\bibinfo{collaboration}{CP-PACS}), \bibinfo{journal}{Phys. Rev.}
  \textbf{\bibinfo{volume}{D65}}, \bibinfo{pages}{094508}
  (\bibinfo{year}{2002}), \eprint{hep-lat/0112020}.

\bibitem[{\citenamefont{Morningstar and Peardon}(1999)}]{Morningstar:1999rf}
\bibinfo{author}{\bibfnamefont{C.~J.} \bibnamefont{Morningstar}}
  \bibnamefont{and} \bibinfo{author}{\bibfnamefont{M.~J.}
  \bibnamefont{Peardon}}, \bibinfo{journal}{Phys. Rev.}
  \textbf{\bibinfo{volume}{D60}}, \bibinfo{pages}{034509}
  (\bibinfo{year}{1999}), \eprint{hep-lat/9901004}.

\bibitem[{\citenamefont{Lichtl}(2006)}]{Lichtl:2006dt}
\bibinfo{author}{\bibfnamefont{A.~C.} \bibnamefont{Lichtl}}
  (\bibinfo{year}{2006}), \eprint{hep-lat/0609019}.

\bibitem[{\citenamefont{Basak et~al.}(2007)}]{Basak:2007kj}
\bibinfo{author}{\bibfnamefont{S.}~\bibnamefont{Basak}} \bibnamefont{et~al.},
  \bibinfo{journal}{Phys. Rev.} \textbf{\bibinfo{volume}{D76}},
  \bibinfo{pages}{074504} (\bibinfo{year}{2007}), \eprint{arXiv:0709.0008
  [hep-lat]}.

\bibitem[{\citenamefont{Dudek et~al.}(2008)\citenamefont{Dudek, Edwards,
  Mathur, and Richards}}]{Dudek:2007wv}
\bibinfo{author}{\bibfnamefont{J.~J.} \bibnamefont{Dudek}},
  \bibinfo{author}{\bibfnamefont{R.~G.} \bibnamefont{Edwards}},
  \bibinfo{author}{\bibfnamefont{N.}~\bibnamefont{Mathur}}, \bibnamefont{and}
  \bibinfo{author}{\bibfnamefont{D.~G.} \bibnamefont{Richards}},
  \bibinfo{journal}{Phys. Rev.} \textbf{\bibinfo{volume}{D77}},
  \bibinfo{pages}{034501} (\bibinfo{year}{2008}), \eprint{arXiv:0707.4162
  [hep-lat]}.

\bibitem[{\citenamefont{Harada et~al.}(2001)\citenamefont{Harada, Kronfeld,
  Matsufuru, Nakajima, and Onogi}}]{Harada:2001ei}
\bibinfo{author}{\bibfnamefont{J.}~\bibnamefont{Harada}},
  \bibinfo{author}{\bibfnamefont{A.~S.} \bibnamefont{Kronfeld}},
  \bibinfo{author}{\bibfnamefont{H.}~\bibnamefont{Matsufuru}},
  \bibinfo{author}{\bibfnamefont{N.}~\bibnamefont{Nakajima}}, \bibnamefont{and}
  \bibinfo{author}{\bibfnamefont{T.}~\bibnamefont{Onogi}},
  \bibinfo{journal}{Phys. Rev.} \textbf{\bibinfo{volume}{D64}},
  \bibinfo{pages}{074501} (\bibinfo{year}{2001}), \eprint{hep-lat/0103026}.

\bibitem[{\citenamefont{Aoki et~al.}(2003)\citenamefont{Aoki, Kuramashi, and
  Tominaga}}]{Aoki:2001ra}
\bibinfo{author}{\bibfnamefont{S.}~\bibnamefont{Aoki}},
  \bibinfo{author}{\bibfnamefont{Y.}~\bibnamefont{Kuramashi}},
  \bibnamefont{and} \bibinfo{author}{\bibfnamefont{S.-i.}
  \bibnamefont{Tominaga}}, \bibinfo{journal}{Prog. Theor. Phys.}
  \textbf{\bibinfo{volume}{109}}, \bibinfo{pages}{383} (\bibinfo{year}{2003}),
  \eprint{hep-lat/0107009}.

\bibitem[{\citenamefont{Hashimoto and Okamoto}(2003)}]{Hashimoto:2003fs}
\bibinfo{author}{\bibfnamefont{S.}~\bibnamefont{Hashimoto}} \bibnamefont{and}
  \bibinfo{author}{\bibfnamefont{M.}~\bibnamefont{Okamoto}},
  \bibinfo{journal}{Phys. Rev.} \textbf{\bibinfo{volume}{D67}},
  \bibinfo{pages}{114503} (\bibinfo{year}{2003}), \eprint{hep-lat/0302012}.

\bibitem[{\citenamefont{Umeda et~al.}(2003)}]{Umeda:2003pj}
\bibinfo{author}{\bibfnamefont{T.}~\bibnamefont{Umeda}} \bibnamefont{et~al.}
  (\bibinfo{collaboration}{CP-PACS}), \bibinfo{journal}{Phys. Rev.}
  \textbf{\bibinfo{volume}{D68}}, \bibinfo{pages}{034503}
  (\bibinfo{year}{2003}), \eprint{hep-lat/0302024}.

\bibitem[{\citenamefont{Morrin et~al.}(2006)\citenamefont{Morrin, Cais,
  Peardon, Ryan, and Skullerud}}]{Morrin:2006tf}
\bibinfo{author}{\bibfnamefont{R.}~\bibnamefont{Morrin}},
  \bibinfo{author}{\bibfnamefont{A.~O.} \bibnamefont{Cais}},
  \bibinfo{author}{\bibfnamefont{M.}~\bibnamefont{Peardon}},
  \bibinfo{author}{\bibfnamefont{S.~M.} \bibnamefont{Ryan}}, \bibnamefont{and}
  \bibinfo{author}{\bibfnamefont{J.-I.} \bibnamefont{Skullerud}},
  \bibinfo{journal}{Phys. Rev.} \textbf{\bibinfo{volume}{D74}},
  \bibinfo{pages}{014505} (\bibinfo{year}{2006}), \eprint{hep-lat/0604021}.

\bibitem[{\citenamefont{Sheikholeslami and
  Wohlert}(1985)}]{Sheikholeslami:1985ij}
\bibinfo{author}{\bibfnamefont{B.}~\bibnamefont{Sheikholeslami}}
  \bibnamefont{and} \bibinfo{author}{\bibfnamefont{R.}~\bibnamefont{Wohlert}},
  \bibinfo{journal}{Nucl. Phys.} \textbf{\bibinfo{volume}{B259}},
  \bibinfo{pages}{572} (\bibinfo{year}{1985}).

\bibitem[{\citenamefont{Morningstar and Peardon}(2004)}]{Morningstar:2003gk}
\bibinfo{author}{\bibfnamefont{C.}~\bibnamefont{Morningstar}} \bibnamefont{and}
  \bibinfo{author}{\bibfnamefont{M.~J.} \bibnamefont{Peardon}},
  \bibinfo{journal}{Phys. Rev.} \textbf{\bibinfo{volume}{D69}},
  \bibinfo{pages}{054501} (\bibinfo{year}{2004}), \eprint{hep-lat/0311018}.

\bibitem[{\citenamefont{Luscher et~al.}(1992)\citenamefont{Luscher, Narayanan,
  Weisz, and Wolff}}]{Luscher:1992an}
\bibinfo{author}{\bibfnamefont{M.}~\bibnamefont{Luscher}},
  \bibinfo{author}{\bibfnamefont{R.}~\bibnamefont{Narayanan}},
  \bibinfo{author}{\bibfnamefont{P.}~\bibnamefont{Weisz}}, \bibnamefont{and}
  \bibinfo{author}{\bibfnamefont{U.}~\bibnamefont{Wolff}},
  \bibinfo{journal}{Nucl. Phys.} \textbf{\bibinfo{volume}{B384}},
  \bibinfo{pages}{168} (\bibinfo{year}{1992}), \eprint{hep-lat/9207009}.

\bibitem[{\citenamefont{Luscher et~al.}(1996)\citenamefont{Luscher, Sint,
  Sommer, and Weisz}}]{Luscher:1996sc}
\bibinfo{author}{\bibfnamefont{M.}~\bibnamefont{Luscher}},
  \bibinfo{author}{\bibfnamefont{S.}~\bibnamefont{Sint}},
  \bibinfo{author}{\bibfnamefont{R.}~\bibnamefont{Sommer}}, \bibnamefont{and}
  \bibinfo{author}{\bibfnamefont{P.}~\bibnamefont{Weisz}},
  \bibinfo{journal}{Nucl. Phys.} \textbf{\bibinfo{volume}{B478}},
  \bibinfo{pages}{365} (\bibinfo{year}{1996}), \eprint{hep-lat/9605038}.

\bibitem[{\citenamefont{Luscher et~al.}(1997)\citenamefont{Luscher, Sint,
  Sommer, Weisz, and Wolff}}]{Luscher:1996ug}
\bibinfo{author}{\bibfnamefont{M.}~\bibnamefont{Luscher}},
  \bibinfo{author}{\bibfnamefont{S.}~\bibnamefont{Sint}},
  \bibinfo{author}{\bibfnamefont{R.}~\bibnamefont{Sommer}},
  \bibinfo{author}{\bibfnamefont{P.}~\bibnamefont{Weisz}}, \bibnamefont{and}
  \bibinfo{author}{\bibfnamefont{U.}~\bibnamefont{Wolff}},
  \bibinfo{journal}{Nucl. Phys.} \textbf{\bibinfo{volume}{B491}},
  \bibinfo{pages}{323} (\bibinfo{year}{1997}), \eprint{hep-lat/9609035}.

\bibitem[{\citenamefont{Klassen}(1998{\natexlab{a}})}]{Klassen:1997jf}
\bibinfo{author}{\bibfnamefont{T.~R.} \bibnamefont{Klassen}},
  \bibinfo{journal}{Nucl. Phys.} \textbf{\bibinfo{volume}{B509}},
  \bibinfo{pages}{391} (\bibinfo{year}{1998}{\natexlab{a}}),
  \eprint{hep-lat/9705025}.

\bibitem[{\citenamefont{Edwards and Joo}(2005)}]{Edwards:2004sx}
\bibinfo{author}{\bibfnamefont{R.~G.} \bibnamefont{Edwards}} \bibnamefont{and}
  \bibinfo{author}{\bibfnamefont{B.}~\bibnamefont{Joo}}
  (\bibinfo{collaboration}{SciDAC}), \bibinfo{journal}{Nucl. Phys. Proc.
  Suppl.} \textbf{\bibinfo{volume}{140}}, \bibinfo{pages}{832}
  (\bibinfo{year}{2005}), \eprint{hep-lat/0409003}.

\bibitem[{\citenamefont{Lin et~al.}(2007)\citenamefont{Lin, Edwards, and
  Joo}}]{Lin:2007yf}
\bibinfo{author}{\bibfnamefont{H.-W.} \bibnamefont{Lin}},
  \bibinfo{author}{\bibfnamefont{R.~G.} \bibnamefont{Edwards}},
  \bibnamefont{and} \bibinfo{author}{\bibfnamefont{B.}~\bibnamefont{Joo}}
  (\bibinfo{year}{2007}), \eprint{arXiv:0709.4680 [hep-lat]}.

\bibitem[{\citenamefont{Symanzik}(1983{\natexlab{a}})}]{Symanzik:1983dc}
\bibinfo{author}{\bibfnamefont{K.}~\bibnamefont{Symanzik}},
  \bibinfo{journal}{Nucl. Phys.} \textbf{\bibinfo{volume}{B226}},
  \bibinfo{pages}{187} (\bibinfo{year}{1983}{\natexlab{a}}).

\bibitem[{\citenamefont{Symanzik}(1983{\natexlab{b}})}]{Symanzik:1983gh}
\bibinfo{author}{\bibfnamefont{K.}~\bibnamefont{Symanzik}},
  \bibinfo{journal}{Nucl. Phys.} \textbf{\bibinfo{volume}{B226}},
  \bibinfo{pages}{205} (\bibinfo{year}{1983}{\natexlab{b}}).

\bibitem[{\citenamefont{DeGrand et~al.}(1998)\citenamefont{DeGrand, Hasenfratz,
  and Kovacs}}]{DeGrand:1998jq}
\bibinfo{author}{\bibfnamefont{T.~A.} \bibnamefont{DeGrand}},
  \bibinfo{author}{\bibfnamefont{A.}~\bibnamefont{Hasenfratz}},
  \bibnamefont{and} \bibinfo{author}{\bibfnamefont{T.~G.} \bibnamefont{Kovacs}}
  (\bibinfo{collaboration}{MILC}) (\bibinfo{year}{1998}),
  \eprint{hep-lat/9807002}.

\bibitem[{\citenamefont{Edwards et~al.}(1998)\citenamefont{Edwards, Heller, and
  Klassen}}]{Edwards:1997nh}
\bibinfo{author}{\bibfnamefont{R.~G.} \bibnamefont{Edwards}},
  \bibinfo{author}{\bibfnamefont{U.~M.} \bibnamefont{Heller}},
  \bibnamefont{and} \bibinfo{author}{\bibfnamefont{T.~R.}
  \bibnamefont{Klassen}}, \bibinfo{journal}{Phys. Rev. Lett.}
  \textbf{\bibinfo{volume}{80}}, \bibinfo{pages}{3448} (\bibinfo{year}{1998}),
  \eprint{hep-lat/9711052}.

\bibitem[{\citenamefont{Jansen and Sommer}(1998)}]{Jansen:1998mx}
\bibinfo{author}{\bibfnamefont{K.}~\bibnamefont{Jansen}} \bibnamefont{and}
  \bibinfo{author}{\bibfnamefont{R.}~\bibnamefont{Sommer}}
  (\bibinfo{collaboration}{ALPHA}), \bibinfo{journal}{Nucl. Phys.}
  \textbf{\bibinfo{volume}{B530}}, \bibinfo{pages}{185} (\bibinfo{year}{1998}),
  \eprint{hep-lat/9803017}.

\bibitem[{\citenamefont{Yamada et~al.}(2005)}]{Yamada:2004ja}
\bibinfo{author}{\bibfnamefont{N.}~\bibnamefont{Yamada}} \bibnamefont{et~al.}
  (\bibinfo{collaboration}{JLQCD}), \bibinfo{journal}{Phys. Rev.}
  \textbf{\bibinfo{volume}{D71}}, \bibinfo{pages}{054505}
  (\bibinfo{year}{2005}), \eprint{hep-lat/0406028}.

\bibitem[{\citenamefont{Aoki et~al.}(2006)}]{Aoki:2005et}
\bibinfo{author}{\bibfnamefont{S.}~\bibnamefont{Aoki}} \bibnamefont{et~al.}
  (\bibinfo{collaboration}{CP-PACS}), \bibinfo{journal}{Phys. Rev.}
  \textbf{\bibinfo{volume}{D73}}, \bibinfo{pages}{034501}
  (\bibinfo{year}{2006}), \eprint{hep-lat/0508031}.

\bibitem[{\citenamefont{Clark}(2006)}]{Clark:2006wq}
\bibinfo{author}{\bibfnamefont{M.~A.} \bibnamefont{Clark}}
  (\bibinfo{year}{2006}), \eprint{hep-lat/0610048}.

\bibitem[{\citenamefont{Kennedy}(2006)}]{Kennedy:2006ax}
\bibinfo{author}{\bibfnamefont{A.~D.} \bibnamefont{Kennedy}}
  (\bibinfo{year}{2006}), \eprint{hep-lat/0607038}.

\bibitem[{\citenamefont{Clark and Kennedy}(2005)}]{Clark:2004cq}
\bibinfo{author}{\bibfnamefont{M.~A.} \bibnamefont{Clark}} \bibnamefont{and}
  \bibinfo{author}{\bibfnamefont{A.~D.} \bibnamefont{Kennedy}},
  \bibinfo{journal}{Nucl. Phys. Proc. Suppl.} \textbf{\bibinfo{volume}{140}},
  \bibinfo{pages}{838} (\bibinfo{year}{2005}), \eprint{hep-lat/0409134}.

\bibitem[{\citenamefont{Jegerlehner}(1996)}]{Jegerlehner:1996pm}
\bibinfo{author}{\bibfnamefont{B.}~\bibnamefont{Jegerlehner}}
  (\bibinfo{year}{1996}), \eprint{hep-lat/9612014}.

\bibitem[{\citenamefont{Omelyan and Folk}(2003)}]{Omelyan:2003}
\bibinfo{author}{\bibfnamefont{I.~M.} \bibnamefont{Omelyan},
  \bibfnamefont{I.~P.~Mryglod}} \bibnamefont{and}
  \bibinfo{author}{\bibfnamefont{R.}~\bibnamefont{Folk}},
  \bibinfo{journal}{Comp. Phys. Comm.} \textbf{\bibinfo{volume}{151}},
  \bibinfo{pages}{272} (\bibinfo{year}{2003}).

\bibitem[{\citenamefont{de~Forcrand and Takaishi}(1997)}]{deForcrand:1996ck}
\bibinfo{author}{\bibfnamefont{P.}~\bibnamefont{de~Forcrand}} \bibnamefont{and}
  \bibinfo{author}{\bibfnamefont{T.}~\bibnamefont{Takaishi}},
  \bibinfo{journal}{Nucl. Phys. Proc. Suppl.} \textbf{\bibinfo{volume}{53}},
  \bibinfo{pages}{968} (\bibinfo{year}{1997}), \eprint{hep-lat/9608093}.

\bibitem[{\citenamefont{Sexton and Weingarten}(1992)}]{Sexton:1992nu}
\bibinfo{author}{\bibfnamefont{J.~C.} \bibnamefont{Sexton}} \bibnamefont{and}
  \bibinfo{author}{\bibfnamefont{D.~H.} \bibnamefont{Weingarten}},
  \bibinfo{journal}{Nucl. Phys.} \textbf{\bibinfo{volume}{B380}},
  \bibinfo{pages}{665} (\bibinfo{year}{1992}).

\bibitem[{\citenamefont{Weingarten and Petcher}(1981)}]{Weingarten:1980hx}
\bibinfo{author}{\bibfnamefont{D.~H.} \bibnamefont{Weingarten}}
  \bibnamefont{and} \bibinfo{author}{\bibfnamefont{D.~N.}
  \bibnamefont{Petcher}}, \bibinfo{journal}{Phys. Lett.}
  \textbf{\bibinfo{volume}{B99}}, \bibinfo{pages}{333} (\bibinfo{year}{1981}).

\bibitem[{\citenamefont{Hasenbusch and Jansen}(2003)}]{Hasenbusch:2002ai}
\bibinfo{author}{\bibfnamefont{M.}~\bibnamefont{Hasenbusch}} \bibnamefont{and}
  \bibinfo{author}{\bibfnamefont{K.}~\bibnamefont{Jansen}},
  \bibinfo{journal}{Nucl. Phys.} \textbf{\bibinfo{volume}{B659}},
  \bibinfo{pages}{299} (\bibinfo{year}{2003}), \eprint{hep-lat/0211042}.

\bibitem[{\citenamefont{Sommer}(1994)}]{Sommer:1993ce}
\bibinfo{author}{\bibfnamefont{R.}~\bibnamefont{Sommer}},
  \bibinfo{journal}{Nucl. Phys.} \textbf{\bibinfo{volume}{B411}},
  \bibinfo{pages}{839} (\bibinfo{year}{1994}), \eprint{hep-lat/9310022}.

\bibitem[{\citenamefont{Edwards et~al.}(2008)\citenamefont{Edwards, Joo, Lin,
  and Peardon}}]{Edwards:aniso_tune}
\bibinfo{author}{\bibfnamefont{R.~G.} \bibnamefont{Edwards}},
  \bibinfo{author}{\bibfnamefont{B.}~\bibnamefont{Joo}},
  \bibinfo{author}{\bibfnamefont{H.-W.} \bibnamefont{Lin}}, \bibnamefont{and}
  \bibinfo{author}{\bibfnamefont{M.~J.} \bibnamefont{Peardon}},
  \bibinfo{journal}{Work in progress}  (\bibinfo{year}{2008}).

\bibitem[{\citenamefont{Lepage and Mackenzie}(1993)}]{Lepage:1992xa}
\bibinfo{author}{\bibfnamefont{G.~P.} \bibnamefont{Lepage}} \bibnamefont{and}
  \bibinfo{author}{\bibfnamefont{P.~B.} \bibnamefont{Mackenzie}},
  \bibinfo{journal}{Phys. Rev.} \textbf{\bibinfo{volume}{D48}},
  \bibinfo{pages}{2250} (\bibinfo{year}{1993}), \eprint{hep-lat/9209022}.

\bibitem[{\citenamefont{Foley}(2008)}]{Foley}
\bibinfo{author}{\bibfnamefont{J.}~\bibnamefont{Foley}},
  \bibinfo{journal}{Private communication}  (\bibinfo{year}{2008}).

\bibitem[{\citenamefont{Capitani et~al.}(2006)\citenamefont{Capitani, Durr, and
  Hoelbling}}]{Capitani:2006ni}
\bibinfo{author}{\bibfnamefont{S.}~\bibnamefont{Capitani}},
  \bibinfo{author}{\bibfnamefont{S.}~\bibnamefont{Durr}}, \bibnamefont{and}
  \bibinfo{author}{\bibfnamefont{C.}~\bibnamefont{Hoelbling}},
  \bibinfo{journal}{JHEP} \textbf{\bibinfo{volume}{11}}, \bibinfo{pages}{028}
  (\bibinfo{year}{2006}), \eprint{hep-lat/0607006}.

\bibitem[{\citenamefont{Klassen}(1998{\natexlab{b}})}]{Klassen:1998ua}
\bibinfo{author}{\bibfnamefont{T.~R.} \bibnamefont{Klassen}},
  \bibinfo{journal}{Nucl. Phys.} \textbf{\bibinfo{volume}{B533}},
  \bibinfo{pages}{557} (\bibinfo{year}{1998}{\natexlab{b}}),
  \eprint{hep-lat/9803010}.

\bibitem[{\citenamefont{Guagnelli et~al.}(1999)\citenamefont{Guagnelli,
  Heitger, Sommer, and Wittig}}]{Guagnelli:1999zf}
\bibinfo{author}{\bibfnamefont{M.}~\bibnamefont{Guagnelli}},
  \bibinfo{author}{\bibfnamefont{J.}~\bibnamefont{Heitger}},
  \bibinfo{author}{\bibfnamefont{R.}~\bibnamefont{Sommer}}, \bibnamefont{and}
  \bibinfo{author}{\bibfnamefont{H.}~\bibnamefont{Wittig}}
  (\bibinfo{collaboration}{ALPHA}), \bibinfo{journal}{Nucl. Phys.}
  \textbf{\bibinfo{volume}{B560}}, \bibinfo{pages}{465} (\bibinfo{year}{1999}),
  \eprint{hep-lat/9903040}.

\end{thebibliography}
\end{document}